# Numerical simulations of the latest caldera-forming eruption of Okmok volcano, Alaska

Alain BURGISSER [1*], Ally PECCIA [2], Terry PLANK [2], Yves MOUSSALLAM [2]

[1] Univ. Grenoble Alpes, Univ. Savoie Mont Blanc, CNRS, IRD, Univ. Gustave Eiffel, ISTerre, 38000 Grenoble, France.

[2] Lamont-Doherty Earth Observatory, Palisades, NY 10964, USA.

* Corresponding author: Phone: (+33) 479 758 705. Email: alain.burgisser@univ-smb.fr

## Abstract

The 2050±50 [14]C yBP caldera-forming eruption of Okmok volcano, Alaska, had a global atmospheric impact with tephra deposits found in distant Arctic ice cores and a sulfate signal found in both Greenland and Antarctic ice cores. The associated global climate cooling was driven by the amount of sulfur injected into the stratosphere during the climactic phase of the eruption. This phase was dominated by pyroclastic density currents, which have complex emplacement dynamics precluding direct estimates of the sulfur stratospheric load. We simulated the dynamics of the climactic phase with the two-phase flow model MFIX-TFM under axisymmetric conditions with several combinations of mass eruption rate, jet water content, vent size, particle size and density, topography, and emission duration. Results suggest that a steady mass eruption rate of $1.2–3.9\times10^{11}$ kg/s is consistent with field observations. Minimal stratospheric injections occur in pulses issued from the central plume initially rising above the caldera center and from successive phoenix ash-clouds caused by the encounter of the pyroclastic density currents with topography. Most of the volcanic gas is injected into the stratosphere by the buoyant liftoff of dilute parts of the currents at the end of the eruption. Overall, 58–64 wt% of the total amount of gas emitted reaches the stratosphere. A fluctuating emission rate or an efficient final liftoff due to seawater interaction are unlikely to have increased this loading. Combined with petrological estimates of the degassed S, our results suggest that the eruption injected 11–20 Tg S into the stratosphere, consistent with the subsequent climate response and Greenland ice sheet deposition. Our results also show that the combination of the source Richardson number and the mass eruption rate is able to characterize the buoyant–collapse transition at Okmok. We extended this result to 141 runs from 10 published numerical studies of eruptive jets and found that this regime diagram is able to capture the first-order layout of the buoyant–collapse transition in all studies except one. An existing multivariate criterion yields the best predictions of this regime transition.

# 1. Introduction

There is a renewed interest in the 43 BC (2050±50 $^{14}$C yBP) caldera-forming eruption of Okmok volcano on the Aleutian island of Umnak, Alaska (Burgisser 2005; Larsen et al. 2007; Larsen et al. 2023; Peccia et al. 2023) because its tephra deposits were found in distant Arctic ice cores and a time-correlative sulfate signal was found in Antarctic ice cores (McConnell et al. 2020; Pearson et al. 2022). This discovery had far-reaching implications as the radiative forcing of this so-called Okmok II eruption likely caused anomalously cold years in the Northern Hemisphere (McConnell et al. 2020; Peccia et al. 2023), inducing inclement weather, famine, and civil unrest in the declining Roman Republic and Ptolemaic Kingdom. There is a wide range of eruptions associated with major climate change during the Holocene, and combining petrological, field, modelling data with other paleoenvironmental/climatic records of these eruptions is key to understanding their impact on environments and civilizations (e.g., Lavigne et al. 2013; Vidal et al. 2016; Peccia et al. 2023).

Large-scale climate cooling is driven by how much sulfur is injected into the stratosphere. The climactic phase of the Okmok II eruption, however, deposited material from pyroclastic density currents (PDCs) that hug topography and covered ~1000 km$^2$ of Umnak and the neighboring island of Unalaska (Fig. 1A). Although impressive at the regional scale (Okmok II is a VEI 6, caldera-forming ignimbrite of intermediate size, Giordano and Cas 2021), the reach of such collapsing eruptive jet is not expected to attain the high altitude required for stratospheric injection. Most of the associated volcanic sulfur, in fact, is expected to have mixed with the local atmosphere, well below the tropopause. This is because any simultaneous buoyant plume of significance would have left recognizable fall deposits interbedded with (or atop) the massive PDC deposits; this is not seen in the field (Burgisser 2005).

Quantitative estimates of the sulfur stratospheric load of ancient eruptions are unfortunately challenging. Defining robust criteria capturing the regime change between buoyant plume and full eruptive column collapse is a long-standing issue. Over the past 33 years, no less than 13 quantities have been proposed to influence this buoyant–collapse transition: mass eruption rate (Bursik and Woods 1991; Costa et al. 2018; Trolese et al. 2019), jet water content (Esposti Ongaro et al. 2006), vent area (Dufek and Bergantz 2007b; Trolese et al. 2019; Esposti Ongaro et al. 2020), jet speed at the vent (Bursik and Woods 1991; Trolese et al. 2019), the product of jet particle concentration and jet speed (Dufek and Bergantz 2007b), temperature (Trolese et al. 2019), jet mixture density (Trolese et al. 2019), caldera filling with pre-existing deposits (Valentine and Cole 2021), and the Rouse number (Valentine and Wohletz 1989). Valentine and Wohletz (1989) have proposed the combination of the Richardson number, the ratio of jet over buoyancy forces, and the ratio of jet over atmospheric pressures. More recently, Koyaguchi and Suzuki (2018) have proposed another multivariate combination: the normalized ratio of the area of the crater top over the crater base, the normalized magma supply rate at the crater base, and the normalized mass eruption rate.

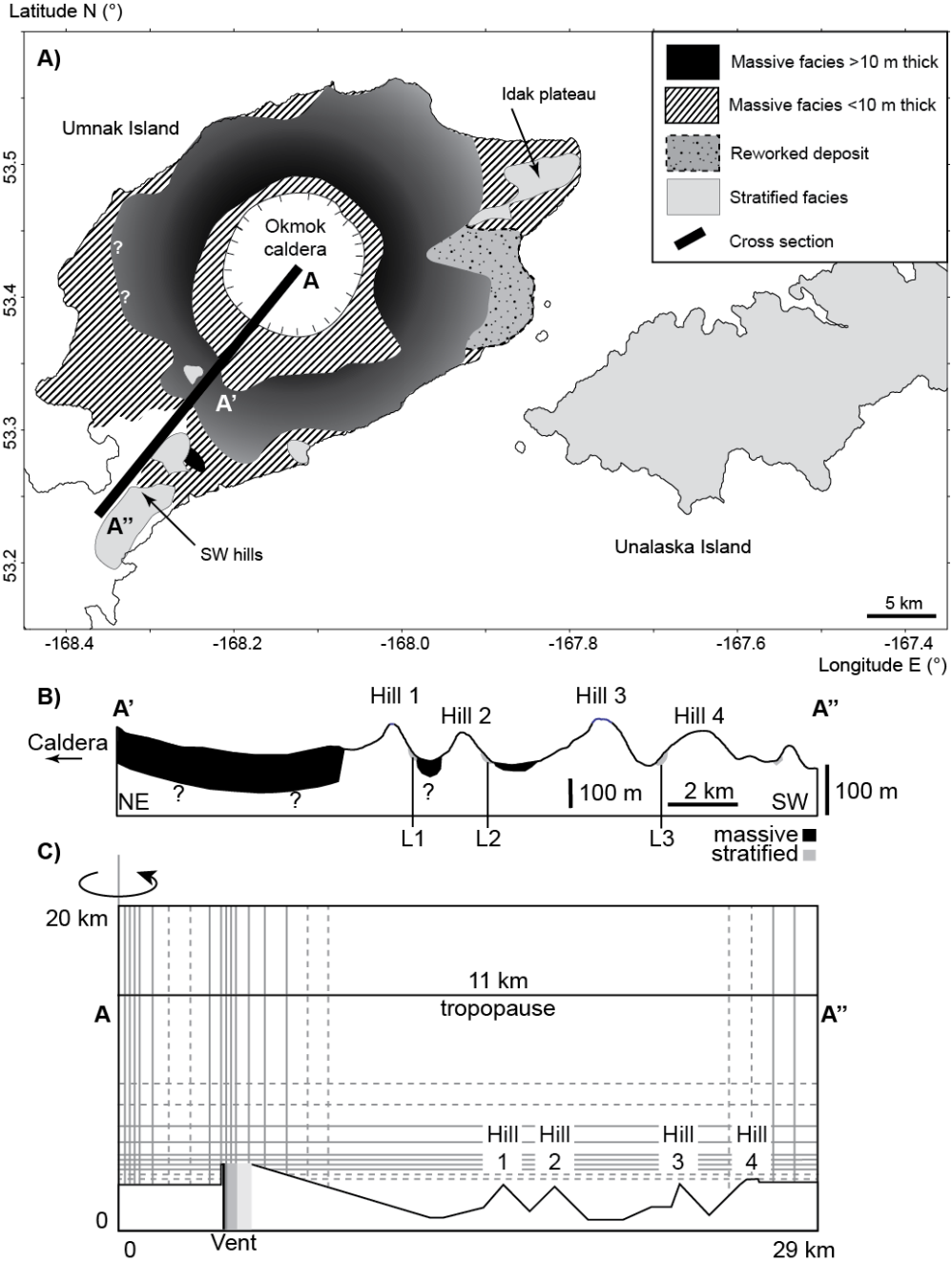

**Figure 1**: Study location and simulation setup. A) Distribution of the pyroclastic deposits of the Okmok II eruption classified by broad type and thickness (modified from Burgisser, 2005). The line A–A'–A'' indicates the location of the cross section shown in B) and C). B) Cross section along A'–A'' with the distribution of massive and stratified deposits. Labels indicate the successive hills (Hill 1–4) that stopped the PDC as well as sampling locations (L1–3) (modified from Burgisser, 2005). C) Discretization of the axisymmetric computational domain of interest (29 by 20 km) along the A–A'' section. The location and extent of the annular vent inner and outer radii are shown in progressively lighter shades of gray corresponding to vent widths of 100, 200, 600, and 1200 m, respectively. For clarity, the computational cells are not to scale and the topography is exaggerated

vertically.

The dynamics of PDCs are not simple; only first-order scaling, for instance, can be drawn between mass eruption rate and PDC runout (Giordano and Cas 2021; Roche et al. 2021). Co-PDC clouds (phoenix ash-clouds hereafter) that are due to the buoyant liftoff from the most dilute parts of PDCs may bring material and volcanic gases into the stratosphere (e.g., Bursik and Woods, 1991; Dufek and Bergantz, 2007a; Calabrò et al. 2022). Such clouds are generated by the complex process of partial or total gravitational collapse of the ejected material and the subsequent density stratification of the resulting PDC (e.g., Neri et al., 2007; Esposti Ongaro et al., 2020; Valentine, 2020). Estimating which fraction of pyroclasts and gas travel as PDCs and which fraction can be injected into the stratosphere from an ancient eruption is nowadays best addressed by numerical modeling (Calabrò et al. 2022).

Here we use the two-phase flow model MFIX-TFM to simulate the dynamics of the climatic phase of Okmok II eruption. MFIX-TFM has been extensively used to model volcanic flows such as plumes and PDCs (Dufek and Bergantz, 2007a ; Dufek and Bergantz, 2007b ; Dufek et al., 2009; Benage et al., 2014 ; Benage et al., 2016; Sweeney and Valentine, 2017; Valentine and Sweeney, 2018; Breard et al., 2019; Valentine, 2020; Valentine and Cole, 2021). We mostly follow the modeling choices of Dufek and Bergantz (2007a) while incorporating more recent findings (e.g., Breard et al., 2019) to apply MFIX-TFM to the case study of Okmok II. No modeling novelty was attempted, and the dynamics of the PDC travelling over water were ignored.

## 2. Summary of the Okmok II eruption

The depositional sequence of the 2050±50 $^{14}$C yBP (Larsen et al. 2023) Okmok II eruption can be divided into four units of interest to establish links with the ash found in the Greenland ice core. The first unit is a deposit of silicic ash fallout (unit A1 in Burgisser, 2005) with a rhyodacitic composition (Larsen et al. 2007). The second unit is another silicic ash fallout deposit (units A2 and B1-2 in Burgisser, 2005). The third unit is a series of fallout deposits (units C1-3 in Burgisser, 2005) that are all andesitic in composition (Larsen et al., 2007; Peccia et al. 2023) except one sub-layer that is silicic. Finally, the fourth unit is composed of voluminous PDC deposits with a basaltic andesite composition that is more mafic (54-56 wt% $SiO_2$) than the underlying fallout deposits (57-59 wt% $SiO_2$) (Larsen et al. 2007). Stratigraphic observations suggest that the ground was mostly free of snow when the eruption started (Burgisser, 2005). The eruption paused after the emission of the two first units for a period of days to months before resuming with unsteady phreatomagmatic explosions that deposited the third unit. The eruption ended with the climactic, caldera-forming phase of the eruption that deposited the fourth unit.

The most voluminous fallout unit is the first one, but the ash found in the Greenland ice cores is not rhyodacitic; its composition matches the whole rock and matrix glass of the mafic PDC deposits (McConnell et al. 2020; Peccia et al., 2023). Moreover, the presence of sulfur isotopes with mass independent fractionation in the ice core sulfate signal indicates oxidation in a high-ultraviolet environment consistent with travel above the ozone layer in the stratosphere (McConnell et al. 2020). The third unit (and part of the second unit) is of phreatomagmatic origin due to vaporization of the pre-eruptive intracaldera lake (Burgisser, 2005). Despite having a composition that also overlaps with that of the Greenland ice core tephra, this andesitic unit has a small volume (<0.2 $km^3$ Dense Rock Equivalent, DRE), consistent fine granulometry and distribution over both Umnak and Unalaska islands, and corresponding low plume heights (up to 10 separate plumes, each of which did not exceed a few kms in height), which make it a very unlikely candidate for stratospheric injection. The most likely culprit of the stratospheric injection is instead the fourth unit (Fig. 1A), the caldera-forming PDC deposits, which we detail below.

The total volume of material ejected by the basaltic andesitic PDC was at least 14 $km^3$ DRE (Burgisser 2005). The upper limit is far less certain, and a reasonable number based on estimates of the caldera volume (Burgisser 2005) and underwater deposits (Peccia et al. 2023b) yields ~29 $km^3$ DRE, which corresponds to a VEI of 6. The average particle density of the PDC deposits is 1445 $kg/m^3$ with a global componentry of 69 wt% scoria, 26 wt% lithics, and 5 wt% of glass and crystals. The coarsest (≥2 mm) fraction is dominated by scoria, the intermediate fraction is composed of a third of lithics and the rest of scoria, and the finest fraction (≤0.02 mm) is dominated by crystals and glass. The highest water concentration in plagioclase-hosted melt inclusions found in the scoria is 2.34 wt% (Peccia et al. 2023b).

The nearly circular shape of the caldera and deposit componentry suggest that the collapse of the eruptive column caused the PDC to spread axisymmetrically (Burgisser 2005). Proximally, along the N and NW flanks of the caldera, thick accumulations of spatter, bombs, and angular lithic fragments form fines-poor agglutinate deposits that extend up to ~4 km from the caldera rim (Larsen et al. 2007; Larsen et al. 2023). Distally, the maximum PDC runout is unknown over most of the circumference of the island because no underwater data are available. Near the shore, where minor hills promoted sedimentation of stratified deposits, Burgisser (2005) estimated the speed of the parent dilute PDCs by using a kinematic model of transport and sedimentation constrained by granulometry data. Assuming that these estimates represent the average current velocity and that the full volume of on-land deposits was emplaced at that average velocity, Burgisser (2005) calculated a maximum eruption rate of $1.4\times10^{11}$ kg/s.

One exception to the absence of maximum runout data lies on the east, where the PDC crossed an 8-km strait to reach the neighboring island of Unalaska, where it continued its course to a total runout greater than 40 km. Analysis of the stratified deposits covering Unalaska suggests that only the dilute parts of the PDC survived the crossing, which was rapidly and significantly affected by large pumice rafts able to support bouncing of lithics that would otherwise sink (Burgisser 2005). The other exception to the lack of runout constraints lies on the SW of the caldera, where four successive hills located 15–26 km away from the caldera center cause deposit thickness to vanish progressively (Fig. 1B). Thick (up to 30 m), massive PDC deposits were found up to ~22 km from the caldera center, past two successive hills that bore thin, stratified deposits from dilute parts of the PDC on their upper slopes.

Only thin stratified deposits were found further afield, from ~24.5 km to ~29 km in the two shallow valleys separating the last two hills, suggesting deposition from dilute PDC and/or settling phoenix ash-clouds. The total runout along this SW axis is thus at most 29 km away from the caldera center.

## 3. Methods

### *3.1. Model formulation*

We performed numerical simulations by using the MFIX-TFM software (https://mfix.netl.doe.gov, v. 2016), which solves the mass, momentum, and energy balance of a mixture of gas and particles. The gas is modeled as a Newtonian fluid, the turbulence of which is represented by a $k$–$\varepsilon$ model describing the transport of turbulent kinetic energy and its rate of dissipation. Particles are modeled as a one (or three) granular continuum with a given particle size and density. This combination of interpenetrating continuous phases is referred to as the Two-Fluid Model (TFM) approach. Musser and Carney (2020) reviewed the theoretical basis of MFIX-TFM and detailed explanations about the theory and implementation of the model can be found in Garg et al. (2010), Syamlal (1998), Syamlal et al. (1993). Importantly, MFIX-TFM captures the coupling between the gas and the particles as well as inter-particle collisions by using kinetic theory (four-way coupling). Plume and PDC simulations involve significant particle volume fractions, leading to particle-particle collisions and turbulence modulation (Capecelatro and Wagner 2024), which cannot solely be captured by gas–particle and particle–gas coupling (two-way coupling). A validation of the MFIX-TFM approach to model gas-particle flows in flumes has been done by Breard et al. (2019). Although the focus of our study is on the dynamics of the mostly dilute jet, the dynamics of the dense undercurrents and the related phoenix ash-clouds also control stratospheric injection and thus a reasonable degree of accuracy of the propagation of the dense granular flow is needed. The choices of the closure relationships thus closely follow those of Breard et al. (2019), except for the inclusion of turbulence energy dissipation (Dufek and Bergantz, 2007a).

The suspension is composed of a gas phase and $m=1$ to $M$ solid phases (symbols are listed in the Appendix). The conservation of mass and volume is given by:

$$\partial_t(\varepsilon_g \rho_g) + \nabla \cdot (\varepsilon_g \rho_g \boldsymbol{v}_g) = 0 \tag{1a}$$

$$\partial_t(\varepsilon_m \rho_m) + \nabla \cdot (\varepsilon_m \rho_m \boldsymbol{v}_m) = 0 \tag{1b}$$

$$\varepsilon_g + \sum_{m=1}^{M} \varepsilon_m = \varepsilon_g + \varepsilon_s = 1 \tag{1c}$$

where $\varepsilon_g$ is the gas volume fraction, $\varepsilon_s$ is the sum of the solid volume fractions, $\rho_g$ is the gas density, $\boldsymbol{v}_g$ is the gas velocity, and $\varepsilon_m$, $\rho_m$, and $\boldsymbol{v}_m$ are the volume fraction, density, and velocity of particle phase $m$, respectively.

The gas density is a function of an additional scalar advection equation that tracks the amount of injected water vapor within the domain. That scalar, $S$, is used to quantify the mixing with the ambient air so that $\rho_g = S\rho_w + (1-S)\rho_a$, where $\rho_w$ is water density (representing the volcanic gas injected into the atmosphere) and $\rho_a$ is air density, both being obtained from the ideal gas law.

The momentum conservation for the gas and the solids is:

$$\partial_t(\varepsilon_g \rho_g \boldsymbol{v}_g) + \nabla \cdot (\varepsilon_g \rho_g \boldsymbol{v}_g \otimes \boldsymbol{v}_g) + \varepsilon_g \nabla P_g + \nabla \cdot \boldsymbol{S}_g + \sum_M f_d(\boldsymbol{v}_m - \boldsymbol{v}_g) + \varepsilon_g \rho_g \boldsymbol{g} = 0 \tag{2a}$$

$$\partial_t(\varepsilon_m \rho_m \boldsymbol{v}_m) + \nabla \cdot (\varepsilon_m \rho_m \boldsymbol{v}_m \otimes \boldsymbol{v}_m) + \varepsilon_m \nabla P_g + \nabla \cdot \boldsymbol{S}_m - f_d(\boldsymbol{v}_m - \boldsymbol{v}_g) + \varepsilon_m \rho_m \boldsymbol{g} = 0 \tag{2b}$$

where $P_g$ is the gas pressure, $\boldsymbol{g}$ is the gravity acceleration, $\boldsymbol{S}_g$ is the gas stress tensor, $\boldsymbol{S}_m$ is the solid stress tensor, and the drag coefficient, $f_d$, follows the formulation of Gidaspow (1994).

The gas stress tensor, $\boldsymbol{S}_g$, is Newtonian:

$$\boldsymbol{S}_g = -P_g \boldsymbol{I} + \boldsymbol{\tau}_g \tag{3a}$$

$$\boldsymbol{\tau}_g = 2\varepsilon_g \mu_{gt} \left(\boldsymbol{D}_g - \frac{1}{3} tr(\boldsymbol{D}_g)\boldsymbol{I}\right) \tag{3b}$$

$$\boldsymbol{D}_g = \tfrac{1}{2}\left[\nabla\boldsymbol{v}_g + (\nabla\boldsymbol{v}_g)^T\right] \tag{3c}$$

$$\mu_{gt} = min\left(\mu_g + 0.09\rho_g k_t^2 \varepsilon_t^{-1}, 10^3\right) \tag{3d}$$

where $k_t$ is the turbulent kinetic energy and $\varepsilon_t$ is its rate of dissipation. A modified $k$–$\varepsilon$ turbulence model suitable for dilute suspensions is used to calculate $k_t$ and $\varepsilon_t$ (Benyahia et al. 2005). For simplicity (Supplementary Information, SI, Section S1), water vapor and air were assumed to have the same viscosity, $\mu_g$:

$$\mu_g = 1.7 \times 10^{-5} \left(\frac{T_g}{273}\right)^{1.5} \left(\frac{383}{T_g+110}\right) \tag{4}$$

where $T_g$ is the gas temperature. The energy conservation for the gas and the solids reads:

$$\varepsilon_g \rho_g C_{pg}(\partial_t T_g + \boldsymbol{v}_g \cdot \nabla T_g) = \nabla(\kappa_g \nabla T_g) + \sum_M \gamma_m (T_m - T_g) \tag{5a}$$

$$\varepsilon_m \rho_m C_{pm}(\partial_t T_m + \boldsymbol{v}_m \cdot \nabla T_m) = \nabla(\kappa_m \nabla T_m) - \gamma_m (T_m - T_g) \tag{5b}$$

where $C_{pg}$ is the gas heat capacity, $C_{pm}$=1200 J/kg K is the solid heat capacity (Moitra et al. 2020), $T_m$ is the solids temperature, $\gamma_m$ is the interphase heat transfer between the solid phase $m$ and the gas, $\kappa_m$ = 2 W/K m is the solid conductivity (Moitra et al. 2020), and $\kappa_g$ is the gas conductivity given by:

$$\kappa_g = 0.025\left(1 - \sqrt{1-\varepsilon_g}\right)\sqrt{\frac{T_g}{300}} \tag{6}$$

The interphase heat transfer, $\gamma_m$, is given by:

$$\gamma_m = \frac{6\kappa_g \varepsilon_m}{d_m^2} Nu_m \tag{7}$$

where $d_m$ is the particle size of solid phase $m$, $Nu_m$ is the Nusselt number of solid phase $m$, which depends on the gas Prandtl number, on the particle Reynolds number, and on $\varepsilon_g$ (Gunn 1978). The gas heat capacity is:

$$C_{pg} = S C_{pw} + (1 - S) C_{pa} \tag{8}$$

where $C_{pw}$=1810 J/kg K is the water vapor heat capacity and $C_{pa}$=1000 J/kg K is the air heat capacity.

The particle stress tensor, $\boldsymbol{S}_m$, is the sum of a collisional and a frictional contribution, the full formulation of which can be found in Benyahia (2008). Briefly, the collisional component of $\boldsymbol{S}_m$ follows the kinetic theory of granular material with a balance equation of pseudo-thermal energy describing the evolution of the granular temperature (Benyahia et al., 2005; Benyahia, 2008). The particle restitution coefficient is 0.8. The frictional component of $\boldsymbol{S}_m$ is formulated according to the Princeton model (Benyahia, 2008), which has been shown to produce flow front velocities that are in good agreement with experimental gravity currents (Breard et al. 2019). This model calculates a solid pressure based on plastic flow theory. The friction coefficient between particle phases is 0.5 and the angle of internal friction of all particle phases is 30°. The particle volume fraction above which friction sets in is 0.5 and the maximum packing fraction is 0.6. In many runs, the solid pressure calculation was turned off once maximum packing fraction is exceeded (see Section 3.2).

Replacing the modified $k$–$\varepsilon$ turbulence model by a Reynolds stress with a fixed turbulent length scale of 1 cm yielded stratospheric gas velocities in excess of 2 km/s with strong reflections of incoming pressure waves against the upper boundary throughout run duration (700 s). With the $k$–$\varepsilon$ model, such pressure reflections were restricted to the first arrival of the central plume at ~100 s with gas velocities <300 m/s at the upper boundary, where the plume is very dilute and the speed of sound is close to that of a pure gas. Turbulent dissipation had thus to be taken into account to avoid unrealistically high velocities in the upper atmosphere.

## *3.2 Plume model intercomparison*

The study of Suzuki et al. (2016) presents results from four different three-dimensional models of weak (low mass eruption rate, $Q$) and strong (high $Q$) volcanic plumes. We ran 10 simulations in 2D and 3D with MFIX-TFM in the configuration listed in Section 3.1 under the same initial conditions as those used in Suzuki et al. (2016). As detailed in SI Section S1, there is an overall inter-model agreement in plume rise dynamics. For the weak plume, the most notable differences between our model and the four others is that the level of neutral buoyancy is not reached by our runs. This is due to 1) the lack of a term in the gas energy balance (Eq. (5a), see Dartevelle and Valentine, 2007) that accounts for the cooling (heating) due the gas expansion (contraction) during decompression (compression) and 2) enhanced near-vent air entrainment in our runs compared to that of the other models. As a result, our plume shapes are the narrowest of the five models. For the strong plume, which has a mass eruption rate closer to those explored at Okmok, the overall agreement in neutral buoyancy level and plume shape is better, although the entrained air fraction at high altitude is systematically larger in our model than in the other models. In summary, this comparison is not a validation of any of the five models as no outputs were compared to an independent experimental or natural dataset. It nevertheless constrains the 2D grid size necessary to capture the main features of plume collapse dynamics, which we applied to the Okmok simulation setup.

### *3.3 Simulation setup*

The axisymmetric computational domain is depicted in Fig. 1C. Following the grid resolution of Dufek and Bergantz (2007a) and the results of the plume model intercomparison (Section 3.2), it contains 354 cells horizontally and 280 cells vertically with an edge length varying from 10 m near the ground surface, the central axis, and the annular vent to 100 m near the outer boundaries of the domain of interest. A near-topography resolution of 10 m is generally considered as a good compromise to capture collapsing eruptive columns generating PDCs without excessive computational power (Valentine and Cole 2021). The domain of interest is 29 km wide by 20 km high (except runs 6 and 7 that are only 21 km wide). Many runs were conducted on larger domains to capture the large-scale phoenix ash-clouds that occur at the end of the eruption (Table 1). Domain enlargement was done horizontally and/or vertically by progressively wider cells spanning from 100 m at the edge of the domain of interest to 1 km at the outer domain boundary. The SI Section S2 gives more details on the ways such domain extension was carried out.

Atmospheric temperature was initialized according to the U.S. standard atmosphere (National Oceanic and Atmospheric Administration 1976) with the tropopause at 11 km, based on the high latitude of Okmok volcano (53.4°N) and summertime atmospheric conditions (Liu et al. 2014). The fact that the tropopause is a couple of kilometers lower during winter was neglected because there are stratigraphic indications that the climactic part of the eruption occurred before the winter months. The consequences of this choice are probably minimal because large-scale simulations of the stratospheric injection of S from the Okmok II eruption suggest that the eruption season does not alter the climate response (Peccia et al. 2023).

Twenty-four runs were carried out with similar topography but varying the number and abundance of particle phases, mass eruption rate, duration, inlet width, and domain sizes (Table 1). One additional simulation (run 12) included a filled-in caldera. The ground surface topography is taken from a rasterized DEM vertical section of Umnak Island, which has a spatial resolution of 10 m (U.S. Geological Survey EROS Data Center 1999). The cross section follows the DEM from the caldera center towards a series of ~350-m-high hills on the SW of the caldera (Fig. 1A-C) until the top of the fourth hill at 26 km from the caldera center, after which the ground is assumed to be flat until the distal edge of the computation domain at 29–83 km (Fig. 1C). The distance of 29 km corresponds to the position of a fifth hill that marks the maximum runout of the Okmok II PDCs (Fig. 1B). Ignoring the topography of the 26–29 km range avoided creating a trough near the domain edge that would drive artificial vortexes influencing phoenix ash-cloud dynamics. Any PDC reaching the edge of the domain of interest at 29 km was considered to have a runout exceeding that observed.

The boundary condition for the ground surface was that of Jenkins (1992) as implemented into MFIX-TFM by Benyahia et al. (2007). The low friction limit of this condition allows partial slip of the solids phase against the ground by particle sliding. The coefficient of restitution of the particle–ground collision was 0.7, and the angle of internal friction at the walls was set to the low value of 12°. A test run with a no-slip boundary condition with a vanishing granular temperature at the boundary yielded a current head reaching 15% longer runout at a given time compared to the Jenkins condition. As a result, the central plume was 4% lower at that moment. Changes in ground boundary conditions thus do not significantly influence the stratospheric load.

In most cases, the distal vertical boundary was a free outflow/inflow surface (e.g., Valentine and Sweeney, 2018). In some cases (Table 1), this boundary was replaced by a free-slip wall. This condition was chosen whenever instabilities occurred with the free flow boundary, which was often the case when the inlet was shut down to capture the waning phase of the eruption. Such changes in boundary condition have a minor impact on stratospheric gas injection (SI Section S2).

The upper boundary conditions were $P_g$=5475 Pa for 20-km high runs (67 and 0.23 Pa for 51 and 85 km high runs, respectively), $k_t$=0.01 J/kg, $\varepsilon_t$=0.1 W/kg, and a granular temperature of 0.01 m$^2$/s$^2$ when particles are present. The inner edge of the annular vent was located on the caldera rim at 4.3 km from the central axis (Fig. 1C). The vent width was 200 m for all runs except runs 3, 4, 11 and 17, where it varied between 100 and 1200 m (Table 1). Gas and particles were injected vertically at 105 kPa with $k_t$=7.5 J/kg, $\varepsilon_t$=75 W/kg, a granular temperature of 0.1 m$^2$/s$^2$ and at 1300 K, which corresponds to the low range of the inferred temperatures of the basaltic andesite erupted by Okmok volcano in 2008 (Larsen et al. 2013). Injection temperature was not varied as its effect on column collapse and PDC generation is subordinate to that of mass eruption rate (Todesco et al., 2002; Todesco et al., 2006). A range of mass eruption rate spanning behaviors from buoyant plumes to collapsing columns was generated by changing the injection speed, $v_{in}$, from 150 to 900 m/s and the particle concentration, $\varepsilon_s$, from 0.15 to 8 vol% (Table 1). The range of $\varepsilon_s$ values were chosen to span jet water content ranging from 0.1 to 6 wt%. Considering that the deposits contain 74 wt% of juvenile material with a maximum water content of 2.34 wt%, a reasonable upper estimate of jet water content is 0.74×2.34=1.7 wt%, a value shared by several runs (Table 1).

The characteristics of the solid phases were mostly based on the deposit componentry and size distribution. Runs with three solid phases had injections of equal volume fractions (i.e. $\varepsilon_s$ divided by 3,

Table 1) of particles with densities of 2000, 1500, and 2000 kg/m$^3$ and diameters of 0.02, 0.2, and 2 mm, respectively. Assuming, like Burgisser (2005), densities of 1000, 2500, and 2000 kg/m$^3$ for scoria, lithics plus crystals, and glass, respectively, the densities of the 0.02 and 0.2 mm solid phases corresponds to the observed componentry. Although the observed componentry of the coarsest fraction is dominated by scoria, we decided to use a density for the 2-mm solid phase corresponding to a third of scoria and the rest of lithics. For simplicity, runs with a single solid phase had 0.2-mm particles of 2000 kg/m$^3$, except run 16 that had the observed average particle density of 1445 kg/m$^3$. These choices have a negligible influence on column collapse dynamics and on stratospheric loading (SI Section S3).

In some cases, minor adjustments of inlet speed, solid pressure calculations, or granular temperature calculations were necessary to ensure convergence. These adjustments had minor consequences on eruptive dynamics (SI Section S4).

### *3.4 Free parameters and quantifying stratospheric injection*

There are three weakly constrained input parameters that were allowed to vary between reasonable bounds: the outer radius of the annular vent, $r_2$, the injection speed, $v_{in}$, and the particle concentration, $\varepsilon_s$. These parameters directly control boundary conditions, but they were recast in more convenient forms amenable to study intercomparison. The vent outer radius was converted into a vent width according to $W = r_2 - r_1$ with $r_1$=4.3 km. Instead of monitoring injection speed, $v_{in}$, it is convenient and customary to use mass eruption rate:

$$Q = v_{in}\rho_b\pi(r_2^2 - r_1^2) \qquad (9)$$

where $\rho_b = \varepsilon_g\rho_g + \sum_M \varepsilon_m\rho_m$ is the bulk density with $\rho_g=\rho_w$ at the inlet. When gas and particle densities at the inlet are kept constant between runs, there is a direct link between $\varepsilon_s$ and the mass fraction of water, $x_{w(v)}$:

$$x_{w(v)} = \frac{(1-\varepsilon_s)\rho_w}{\rho_b} \qquad (10)$$

As summarized in Fig. 2, runs covered the following range of three input parameter: $0.1 \leq x_{w(v)} \leq 6$ wt%, $100 \leq W \leq 1200$ m, and $3.5 \times 10^9 \leq Q \leq 2 \times 10^{11}$ kg/s. The other dependent variables used to characterize vent conditions for the study intercomparison (Section 5.2) are the vent area ($A$), $\varepsilon_s v_{in}$ (Dufek and Bergantz 2007b), the source Richardson number, $Ri$ (ratio of kinetic energy over buoyancy, Valentine and Wohletz 1989):

$$Ri = \frac{\rho_b v_{in}^2}{(\rho_b - \rho_a) g R_{eq}} \tag{11}$$

where $R_{eq} = \sqrt{A/\pi}$ is the equivalent radius of the vent. The source $Ri$ is thus partly controlled by the mass fraction of water at the inlet. This initial jet water content is not necessarily equal to the total amount of water exsolved at the fragmentation level; it is the amount of that water that is liberated by fragmentation. In other words, $x_{w(v)}$ is the amount of gas propelling the jet. This distinction is used to estimate the total S stratospheric load.

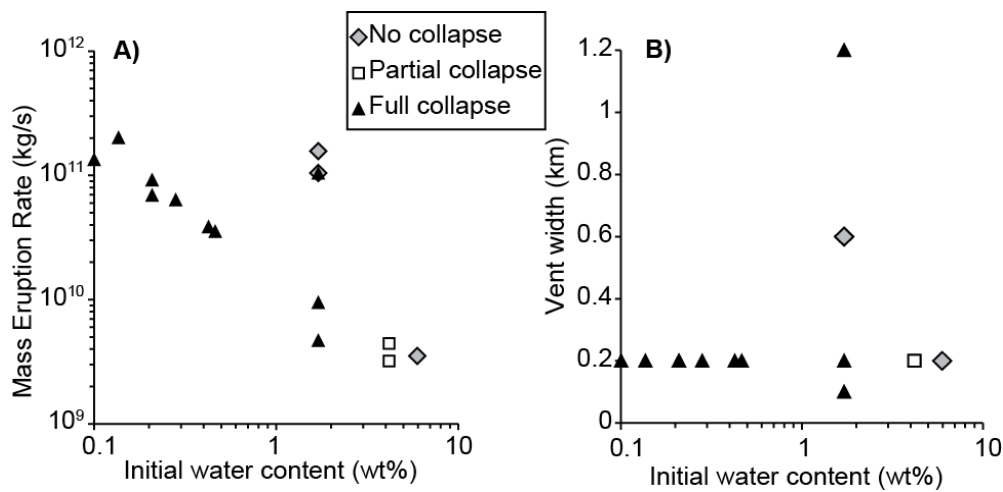

**Figure 2**: Explored values of initial conditions of mass eruption rate, vent width, and initial jet water content (Table 1). Diamonds are runs producing buoyant plumes without collapse, squares are runs producing partial column collapse, and triangles are runs producing full column collapse with PDCs.

In addition, four dimensionless numbers that are defined in SI Section S5: the Mach number, $Ma$, the ratio of the driving force of the jet to the negative buoyancy force of the jet mixture, $Tg_m$, the ratio of vent pressure over atmospheric pressure, $K_p$, and the Rouse number, $P_n$.

Koyaguchi and Suzuki (2018) and Koyaguchi et al. (2018) define three dimensionless parameters ($\tilde{A}$, $\tilde{q}_b$, and $\tilde{Q}$) to discriminate buoyant columns from collapsing columns. The two first parameters are the normalized ratio of the cross-sectional areas at the crater top and base and the normalized magma supply rate at the crater base, respectively:

$$\tilde{A} = Ma^{-1} e^{(Ma^2-1)/2} \qquad (12)$$

$$\tilde{q}_b = K_t^{-1} e^{(Ma^2-1)/2} \quad \text{with} \quad K_t = P_v/P_a = \frac{\rho_b x_{w(v)} R_g T_m}{P_a} \qquad (13)$$

where $T_m$ is the magmatic temperature, $P_v$ is the inlet gas pressure, $P_a$ is the pressure of the atmosphere just above the vent, which we estimated using the air density at the vent heated to magmatic temperature (Koyaguchi and Suzuki (2018) treat the gas–pyroclasts mixture as a single-phase gas). The third parameter is a normalized mass eruption rate:

$$\tilde{Q} = \frac{Q(\rho_b - \rho_a)^2 g^2}{\pi \rho_b^3 Ri_K^2} \left(\frac{Ma}{v_{in}}\right)^5 \qquad (14)$$

where $Ri_K$ is a critical Richardson number defined in SI Section 5. The column is predicted to collapse when

$$\tilde{Q} \geq [Ma + Ma^{-1}(1 - K_t^{-1})]^5 \qquad (15)$$

The efficiency of stratospheric injection was measured by mass balance. The fraction of erupted water mass in the stratosphere is:

$$F_g = \frac{M_{w(out)}}{M_{w(in)}} \qquad (16)$$

where $M_{w(in)}$ and $M_{w(out)}$ are the cumulative masses of water leaving the vent and entering the stratosphere as a function of time, respectively. Computational details are given in SI Section S5.

## 4. Results

The combination of mass eruption rate, vent width, and jet water content can be constrained in three ways. First, simulations should yield plume collapse and fountaining (all of the ejecta collapses back towards the ground and forms PDC; the term *fountaining* being inherited from the resulting shape of the erupting column) that feed PDCs instead of buoyant plumes, as is consistent with the observed deposits. Second, the simulated eruption volume should be ≥14 km$^3$ DRE. The last constraint is the PDC runout of 26–29 km from the caldera center in the SW direction (Fig. 1B). Thus, we deemed runs producing PDCs that reached 26 km after ≥14 km$^3$ DRE of solids were emitted to be compatible with field observations.

## *4.1 Steady mass eruption rate*

Figure 3A shows PDC runouts with fixed topography (vent width of 200 m), steady mass eruption rate, and similar particle densities regardless of size. Under such conditions, our results suggest that the boundary between buoyant plume and fountain collapse is between $4.4\times10^9$ and $9.4\times10^9$ kg/s. A minimum of ~$5\times10^9$ kg/s is thus necessary to account for the thick PDC deposits covering Umnak Island (Fig. 1A). The PDC of the two runs (18 and 19) with the highest mass eruption rates ($1.3–2\times10^{11}$ kg/s) reach 26 km after 14–16 km$^3$ DRE being emitted. Both are thus compatible with field observations, but in different ways. The runout of run 18 remains at 26 km for a few more km$^3$ of erupted material before stabilizing at the apex of hill 3 at ~23 km until the end of the simulation (32 km$^3$ DRE of erupted material). This steady runout position is caused by the buoyant liftoff of the PDC (Bursik and Woods 1996). Run 18 is thus compatible with field observations if the emitted volume is >14 km$^3$ DRE. The PDC of run 19 continues past hill 4 without experiencing buoyant liftoff (a movie is available in SI). Run 19 is thus compatible with field observations if the emitted volume is ~16 km$^3$ DRE.

Although the vent shape is constrained by that of the present caldera, the width of this annular vent is less certain. We chose as a reference the run injecting 1.7 wt% H$_2$O at the vent (run 9, $9.4\times10^9$ kg/s) because it is a reasonable upper estimate of $x_{w(v)}$. We then kept $x_{w(v)}$ constant while varying vent width from the original 200 m to 100, 600, and 1200 m, respectively. For the narrow, 100-m vent, all other parameters were also kept constant, resulting in a smaller mass eruption rate. For the wider vents, the injection speed was increased to match the ~$10^{11}$ kg/s of the 200-m vent runs compatible with field observations.

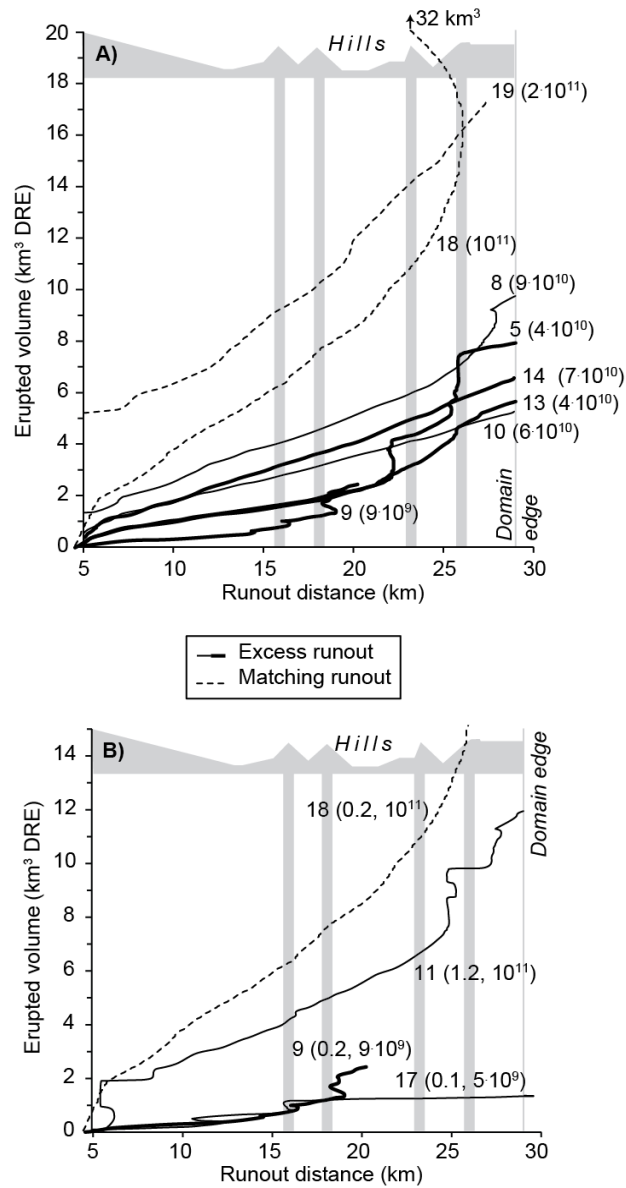

**Figure 3**: PDC runout as a function of erupted volume. Continuous curves represent runs with PDC runout matching that observed at too low an erupted volume (Excess runout) and dashed curves are runs with a plausible combination of runout and erupted volume (Matching runout). Vertical lines mark topographical obstacles, the gray profile is a schematic topography, and the edge of the domain of interest is indicated. A) Vent width of 200 m. Labels indicate run number with the mass eruption rate in parenthesis (kg/s). B) Jet water content of 1.7 wt%. Labels indicate run number with, in parenthesis, the vent width (km) followed by the mass eruption rate (kg/s). Note that the reference run 9 was stopped as soon as it was clear that it would reach the observed runout after an insufficient erupted volume of 4–8 km$^3$ DRE.

Figure 3B shows that the narrower vent with a lower mass eruption rate reaches the maximum observed runout at the insufficient erupted volume of ~1 km$^3$ DRE, which seems to confirm that large mass eruption rates are necessary to erupt enough material before reaching the observed runout. Increasing $v_{in}$ is the only way to increase the erupted volume while matching the observed PDC runout and keeping vent width and $x_{w(v)}$ constant. Such a narrow vent, however, would require $v_{in}$=3.7 km/s to reach 10$^{11}$ kg/s, which corresponds to an unrealistically high Mach number of 37. Vents wider than 100 m can achieve large mass eruption rates at the same water content with a more modest speed increase. This was tested for 600 and 1200 m vents, and the results suggest that mass eruption rate is not the sole control of the buoyant–collapse transition. A vent width of 1200 m at ~10$^{11}$ kg/s reaches the maximum observed runout at only half the needed amount of erupted material (run 11). This run has a much smaller $Ma$ (2.8) and a higher $x_{w(v)}$ (1.7 wt%) than those of the compatible runs with 200-m vents (6–11 and 0.1 wt%, respectively). Vent widths of 600 m, however, produce buoyant plumes despite having similarly high mass flow rates (runs 3 and 4, Table 1). There is thus a need to rationalize the controls of PDC runout, erupted volume, and the buoyant–collapse transition.

Figure 4 shows that PDC propagation speeds change with time in complex ways. Proximal fountain collapse dynamics affects initial runout behavior (<100 s), because larger mass eruption rates cause the collapsing fountain to hit the ground further away from the vent than lower eruption rates. The fountaining jets of the compatible runs 18 and 19 land 4–5 km away from the vent. More distally, the mass eruption rate, $Q$, (and not the speed of the PDC head) strongly controls how the topographic barriers made by the successive hills are overcome. At $Q$ values just above the buoyant–collapse transition, the PDCs of runs 9 and 17 are temporarily blocked by the first two hills. At higher $Q$ values, PDCs are able to jump from the first to the second hill, skipping the valley in between. Although these particular topographic barriers have a limited significance in other settings, they provide an important criterion to understand eruptive dynamics in general. Figure 5 shows that the distal hills have a weak to insignificant effect on runs with high $Q$ (>$4\times10^{10}$ kg/s) and low source Richardson numbers ($Ri$<10).

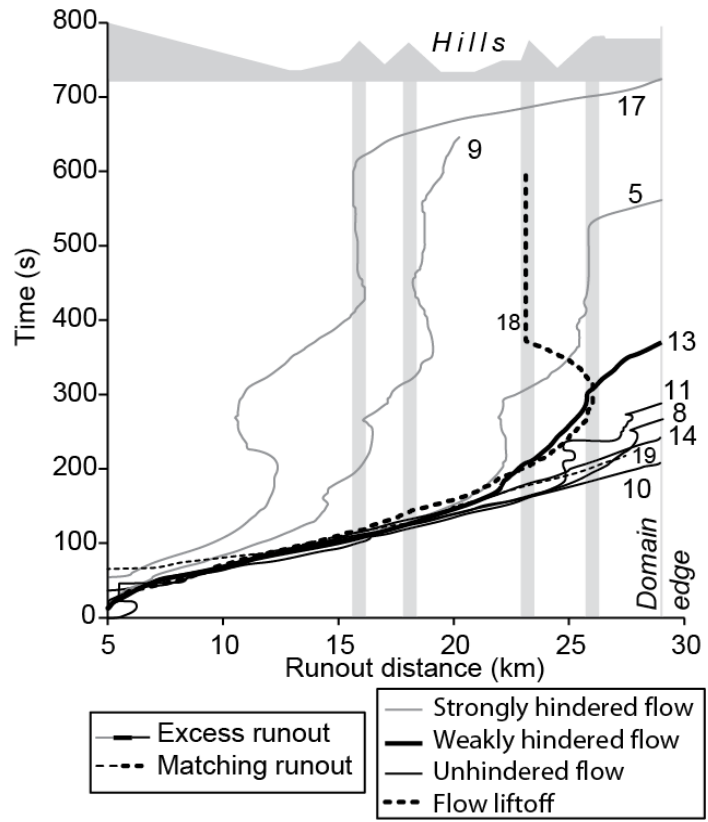

**Figure 4**: Time evolution of the PDC runout for the same runs as in Fig. 3. Continuous curves represent runs with PDC runout matching that observed at too low an erupted volume (Excess runout) and dashed curves are runs with a plausible combination of runout and erupted volume (Matching runout). Runs are classified according to whether PDCs were strongly hindered (thin gray curves), weakly hindered (thick black curve), or unhindered (thin black curves) by topography, or underwent buoyant liftoff (thick dashed curve). Vertical lines mark topographical obstacles, the gray profile is a schematic topography, and the edge of the domain of interest is indicated.

Importantly, the $Q$ vs. $Ri$ regime diagram of Fig. 5 also discriminates the buoyant plume regime ($Ri$>4–7 when $Q$<$10^{10}$ kg/s) from the collapsing regime ($Ri$<14 when $Q$>$10^{10}$ kg/s) in a way that is more robust than any combination of $Q$, $x_{w(v)}$, and $W$ (Fig. 2). This is not surprising as the source Richardson number quantifies the balance between kinetic energy and buoyancy at the vent; a jet with high $Ri$ is more likely to have enough kinetic energy to rise until the initially negative buoyancy of the jet becomes positive. The regime transition is sensitive to small $Q$ variations; a change of -21% to +13% is sufficient to shift from partial collapse to no or full plume collapse, respectively (Fig. 2). This transition is not sensitive to particle density; changing density from 2000 (run 15) to 1445 kg/m³ (run 16) did not induce a regime change and led to broadly similar $F_g$ evolutions (SI Section S6). This is consistent with the modest change in source $Ri$ between these two runs (from 4.4 to 7.4, respectively, SI Table S2).

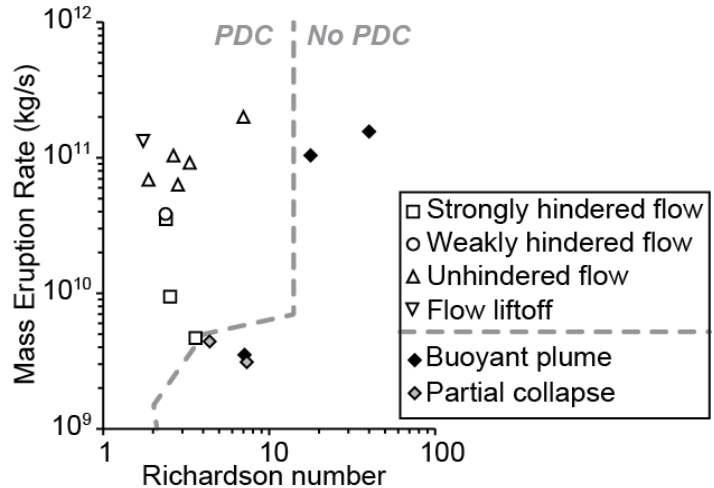

**Figure 5**: Explored values of initial conditions of mass eruption rate and source Richardson number (SI Section S9). Black diamonds are runs producing buoyant plumes without collapse, gray diamonds are runs producing partial column collapse, and open symbols are runs producing full column collapse with PDCs that were strongly hindered (squares), weakly hindered (circles), or unhindered (upturned triangles) by topography, or that underwent buoyant liftoff (downturned triangle). The grey dashed line is the same as that shown in Fig. 13F and it separates fully collapsing columns producing PDCs from buoyant plumes and partially collapsing columns.

The fastest PDCs reaching 26 km are unaffected by topography and do so in ~3 min, whereas PDCs affected by topography may take up to 11 min or more (Fig. 4, SI Section S7). For cases without terminal buoyant liftoff (all runs but 18), we can define a critical mass eruption rate, $Q_{cr}$, at which the observed volume, $V_{obs}$, is emitted when the PDCs reach 26 km as $Q_{cr} = \frac{V_{obs}}{2500 t_{26}}$ kg/s with $t_{26}$=185–646 s. Assuming that PDCs stop propagating as soon as emission stops, the maximum $Q_{cr}$ fulfilling the combined constraints of fountaining eruptions producing unobstructed PDCs with runouts similar to those observed is thus 1.9×10¹¹ kg/s for $V_{obs}$=14 km³ DRE (3.9×10¹¹ kg/s for 29 km³ DRE). If PDCs were temporarily blocked by obstacles, the maximum PDC delaying we observed (run 17) implies a minimum value of $Q_{cr}$ = 5.4×10¹⁰ kg/s.

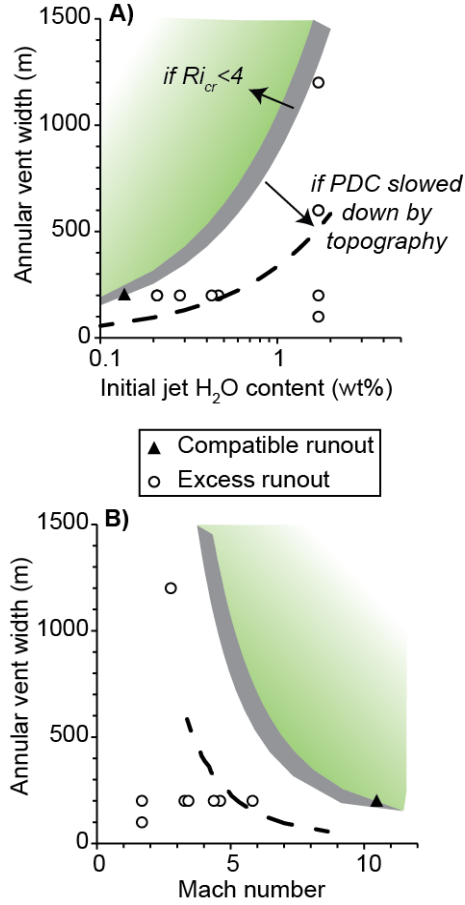

**Figure 6**: Eruption parameters such that the erupted material reaches the volume and extent observed in the field. The green fields cover conditions such that at least 14 km³ DRE of material is erupted when PDCs reach 26 km runout. The thickness of gray curves is defined by $4 \leq Ri_{cr} \leq 7$ and $t_{26}$=185 s ($Q_{cr}$=1.9×10¹¹ kg/s). The triangle is the run compatible with field observations and circles are the other PDC-generating runs without buoyant liftoff. The dashed lines indicate the effect of PDC propagation hindrance by topography ($Ri_{cr}$=7 and $t_{29}$=646 s or $Q_{cr}$=5.4×10¹⁰ kg/s). A) Vent width vs. initial jet water content. B) Vent width vs. Mach number.

Using Eq. (11), a corresponding critical Richardson number can be expressed as:

$$Ri_{cr} = \frac{\pi^{1/2} Q_{cr}^2}{(\rho_b - \rho_a) g \rho_b A^{5/2}} \quad (17)$$

where the vent width can be related to $A$ (Eq. 9) and $x_{w(v)}$ can be related to $\rho_b$ as particle densities vary little in our runs (Eq. 10, SI Section S6). Thus, if $Q_{cr}$ and $Ri_{cr}$ are known, Eq. (17) can be used to calculate the critical vent width above which a volume of $V_{obs}$ has been emitted when the PDCs reach 26 km as a function of $x_{w(v)}$. The jet Mach number can also be evaluated from these quantities. Figure 6 shows such critical vent width with $Ri_{cr}$=4–7 (Fig. 5) and assuming unhindered PDC propagation. To obtain additional runs without buoyant liftoff that are compatible with field observations, one would need to use wide vents with low jet water content and corresponding high *Ma*. To obtain higher water content and lower *Ma*, PDC propagation hindrance by topography is necessary. As explored in Sections 4.2 and 4.3, the natural ejection speed and bulk density are unlikely to be perfectly steady over the eruption duration and some PDCs are subject to propagation inertia after vent shutdown, so this level of precision at a steady eruption rate was deemed sufficient.

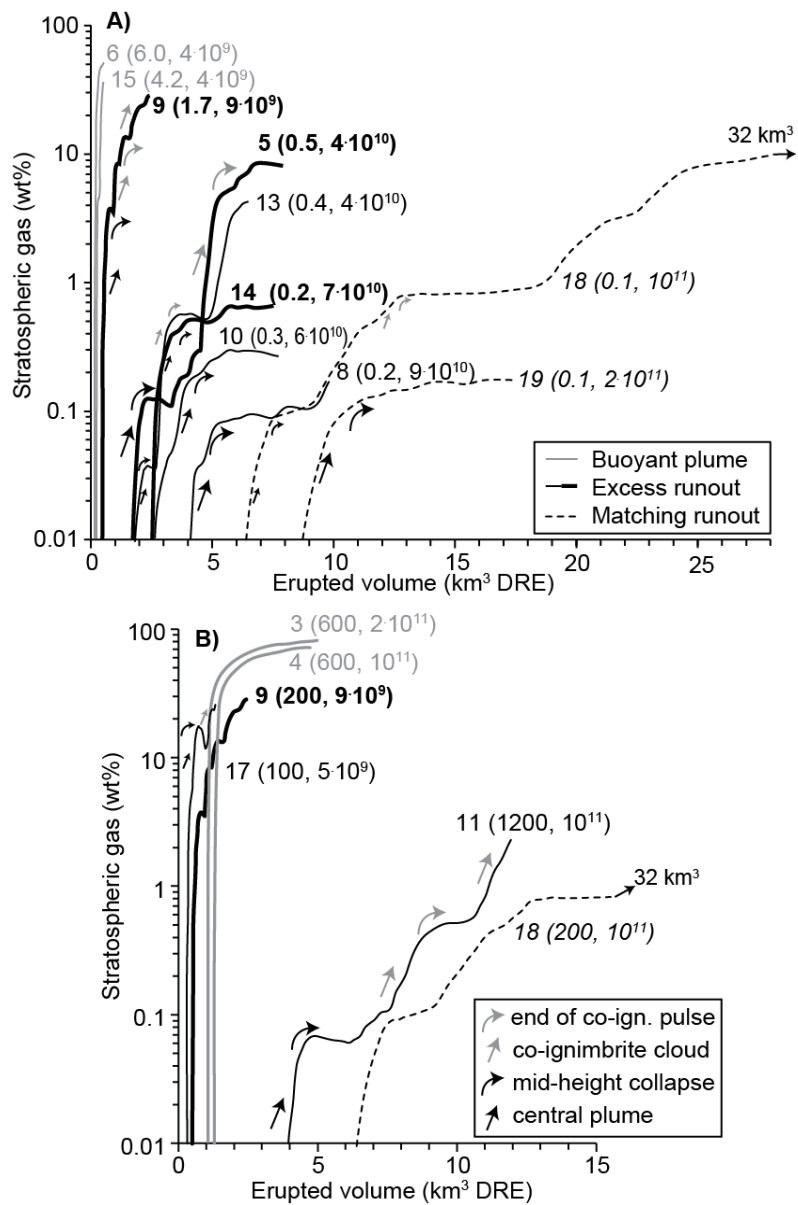

**Figure 7**: Amount of gas injected into the stratosphere ($F_g$) as a function of erupted volume. Gray curves are runs producing buoyant plumes, continuous curves are runs with PDC runout matching that observed at too low an erupted volume (Excess runout) and dashed curves are runs with a plausible combination of runout and erupted volume (Matching runout). A) Vent width of 200 m. Labels indicate

run number with, in parenthesis, the jet water content (wt%) followed by the mass eruption rate (kg/s). Arrows indicate the underlying reason for the curve evolution (see text for more details). B) Jet water content of 1.7 wt%. Labels indicate run number with, in parenthesis, the vent width (m) followed by the mass eruption rate (kg/s).

Figure 7 shows the stratospheric loading for the ten PDC-generating runs and four buoyant runs described above as a function of erupted volume. Leaving the fully buoyant plumes aside (run 6, a movie of which is available in SI, and run 15) to focus on fountaining runs, the first part of the erupted material to reach the stratosphere is systematically the central plume formed by the axial convergence of the pyroclastic material flowing from the ring vent towards the center (Fig. 8). The central plume is not sustained for long, as it collapses back on itself mid-height at an altitude of ~10–11 km, stopping the stratospheric injection. Overall, the central plume injects 0.01–1 wt% of erupted gas above the tropopause. The next injections are caused by the arrival of up to three successive phoenix ash-clouds issued from the main body of the PDCs. These successive clouds inject each an additional few percent of gas above the tropopause, bringing the cumulative amount to <10 wt% at the end of the simulations for all runs except two (Table 1). These two runs (9 and 17, Fig. 7B) are near the buoyant–collapse transition with large excess runouts (Fig. 3) and they inject up to 29 wt% of gas into the stratosphere.

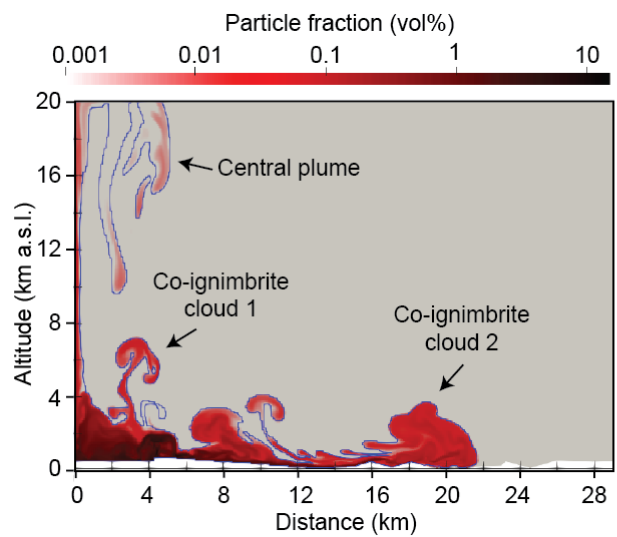

**Figure 8**: Model output example of run 5 at 190 s. The color scale refers to total particle volume fraction. Blue curves outline regions where the gas is above background temperature (>12 °C) and the background atmosphere is shown in gray.

The reasons behind phoenix cloud liftoff are multiple. The first cloud of run 5, for instance, is generated at the apex of the vent because of a favorable configuration of the velocity field generated by air entrainment towards the central plume (Fig. 8, a movie is available in SI). Many clouds are linked to interactions with topography; the second cloud of run 5 is generated by the PDC head encountering hill 2 (Fig. 8). These topographic reliefs cause well-known albeit complex dynamics. The second cloud of run 5 is generated by the PDC going from a supercritical flow regime above hill 2 to a subcritical regime when expanding into the ensuing valley. This incipient phoenix ash-cloud is then fed by more incoming material as the PDC head is nearly stalling because of the deceleration due to this expansion. Another cloud-generating mechanism is due to the fact that some runs (e.g., 8 and 18) emit enough material to fill the caldera with fresh, loose deposits. These deposits are constantly remobilized by the arrival of new material, such that this constant reworking approaches gargle dynamics (Valentine and Cole 2021). The PDC flowing out of the caldera is affected by such gargling because it is sporadically fed by some of dense material overflowing from the caldera. This enhances the generation of small phoenix clouds that merge when ascending towards the stratosphere. In the case of run 18, gargling causes a steady-state situation where these merged clouds feed a steady buoyant liftoff at ~14 km and PDC runout stalls at ~23 km. Assuming a filled caldera with a floor flush with the rim showed that such gargling and the uncertainty in pre-eruptive caldera shape have a limited impact on $F_g$ (SI Section S3).

Overall, stratospheric injection during eruption is modest when column collapse occurs, spanning from 0.1 to <29 wt% (Fig. 7). To rationalize these complex injection patterns, we took all the runs that emitted at least 7 km³ DRE and quantified $F_g$ at that amount of emitted ejecta (Fig. 9). The first-order inverse relationship that can be drawn between $F_g$ and $Q$ confirms that injection of at most a few percent can be expected under compatible conditions. At $Q_{cr}=1.9\times10^{11}$ kg/s, Fig. 9 suggests that <0.01 wt% of volcanic gas would have been injected into the stratosphere when 7 km³ DRE would have been erupted.

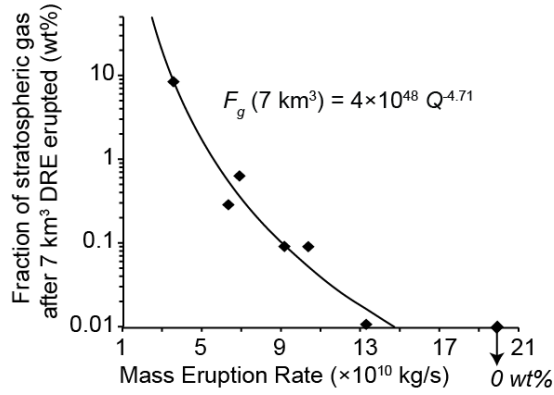

**Figure 9**: Fraction of stratospheric ($F_g$) gas after 7 km³ DRE of ejecta being emitted as a function of mass eruption rate ($Q$) for relevant runs (5, 8, 10, 11, 14, and 18). In addition, run 19 (data point with arrow) has $F_g$≈0 because only a vanishingly small amount of gas has reached the stratosphere by the time that 7 km³ DRE has been emitted. The curve corresponds to the best-fit relationship displayed as a label.

## *4.2 Unsteady mass eruption rate*

Our results show that even minor $Q$ oscillations are likely to straddle the buoyant–collapse transition. Such oscillations could generate a transitory plume during a PDC-dominated eruption, increasing $F_g$. Figure 10A shows the evolution of the mass eruption rate of run 7, which starts in the column collapse regime at $8.6 \times 10^9$ kg/s (close to run 9) for the first 100 s, then switches to the buoyant regime at $3.5 \times 10^9$ kg/s (like run 6) during 200 s before switching back to the column collapse regime (Table 1). Figure 10B compares the evolution of $F_g$ for this unsteady run to the steadily buoyant run 6 and to the fountaining run 9. The transitioning to a purely buoyant plume cause $F_g$ to sharply increases by ~20 wt% above the background fountaining loading by phoenix ash-clouds. The time evolution of $F_g$ (Fig. 10C) indicates that the new buoyant plume emitted at 100 s takes an additional ~100 s to reach the tropopause. The delaying is more straightforward with PDC propagation; the runout stalls for the 200 s duration of the buoyant regime switch (Fig. 10D).

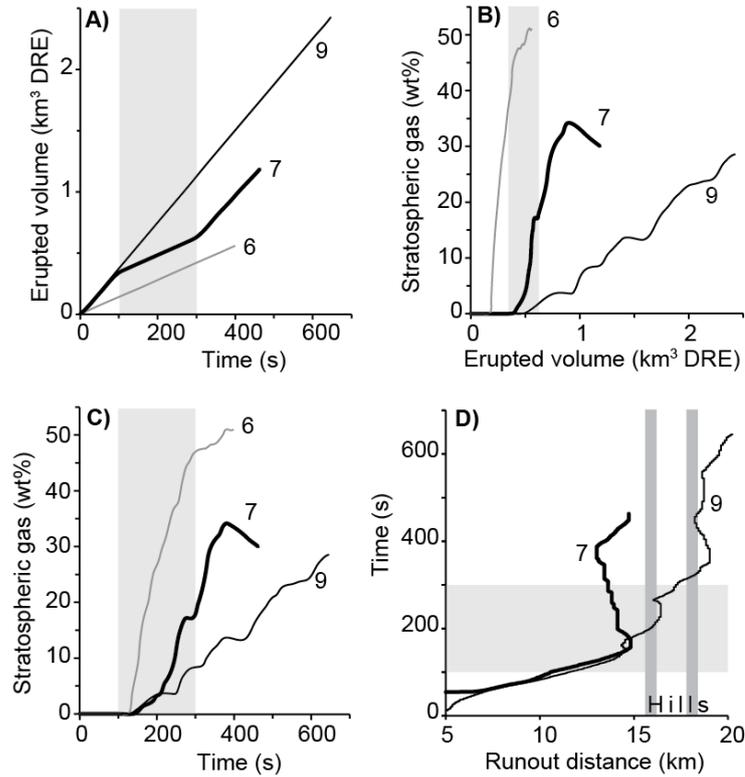

**Figure 10**: Evolution of a run (7, thick black curves) with oscillating mass eruption rate compared to a purely buoyant run (6, gray curves) and a fully collapsing run (9, black curves). The gray areas indicate the 200 s window of the transitory buoyant conditions. A) Time evolution of the erupted volume. B) Amount of gas injected into the stratosphere ($F_g$) as a function of erupted volume. C) Time evolution of $F_g$. D) Time evolution of the PDC runout distance. Vertical lines mark topographical obstacles.

## *4.3 Post eruptive dynamics*

The end of the eruption is likely to send additional volcanic gas at high altitude by liftoff of the most dilute parts of the slowing PDCs. We carried out 8 runs in which emission stopped when PDCs reach the top of the third hill (except runs 17 and 9 where shutdown occurred at hills 1 and 2, respectively). This position was chosen with the intention that PDC head propagation inertia would not exceed a runout of 29 km. Results show, however, that runout inertia cannot be predicted a priori because it broadly follows the complex dynamics of PDCs encounter with topography (SI Section S7), the details of which are beyond the scope of this work. When the eruption stops, Fig. 11 shows that the subsequent liftoff increases $F_g$ in an asymptotic fashion until it reaches a plateau of 58–64 wt%. Some runs, such as run 5, were stopped before reaching this plateau because the large-scale plume dynamics exceeded the size of their computational domain (SI Section S2, a movie is also available in the SI). The amount of gas trapped into the PDC that is being liberated into the stratosphere at the end of the eruption thus systematically exceeds that injected during the eruption.

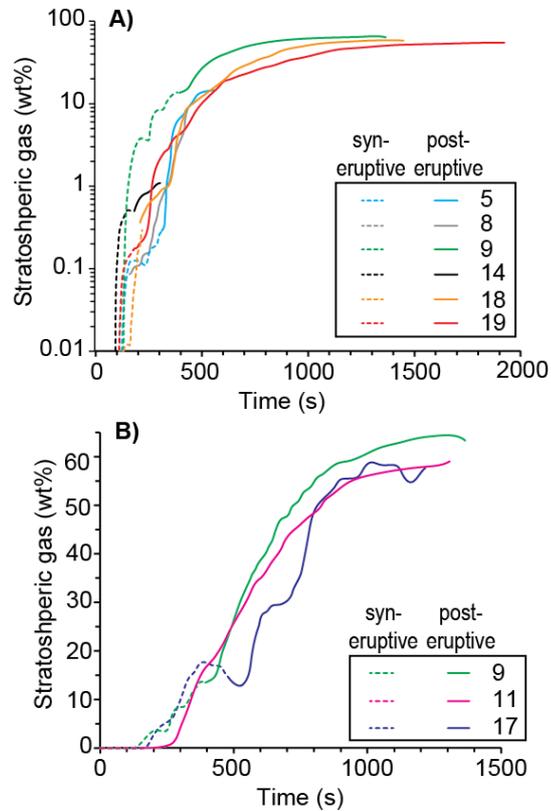

**Figure 11**: Time evolution of stratospheric gas loading ($F_g$). Labels indicate run numbers. The dashed parts of the curves mark the syn-erupting phase and they correspond to the curves on Fig. 7. The solid parts of the curves are the post-eruptive phase. A) Vent width of 200 m. B) Jet water content of 1.7 wt%.

The two runs compatible with observations (runs 18 and 19) only inject 0.1–9 wt% of gas syn-eruptively. They reach post-eruptive plateaus of 55–58 wt% of stratospheric gas (Table 1, a movie of run 19 is available in SI), which is in line with the other measured plateaus. We can thus outline the most likely scenario of eruptive dynamics at high altitude. After a first pulse coming from the central plume, most of the syn-eruptive stratospheric injection is due to a succession of up to three pulses of phoenix cloud material issued from the head and main body of the PDC. The total syn-eruptive injection is thus <1 wt% if no PDC liftoff occurs. The final injection is due to the whole PDC liftoff at the end of the eruption, the dynamics of which seems simply proportional to the erupted volume, yielding a narrow range of $F_g$ values regardless of syn-eruptive PDC steady liftoff. We thus expect a total gas injection of 58–64 wt% into the stratosphere if the mass eruption rate is steady. The tall domain of run 18 ($H_D$=85 km) suggests that most (~90 wt%) of this gas remains into the upper part of the stratosphere because of neutral buoyancy while only a minor amount (~10 wt%) of gas reaches the mesosphere.

## 5. Discussion

Several model limitations need improvement in future work. In the gas energy balance, the lack of a cooling/heating term due to gas decompression/compression (Dartevelle and Valentine, 2007) causes an underestimation of the plume (and phoenix clouds) cooling when reaching the high atmosphere. As a result, neutral buoyancy is bypassed in the weak plume case of the model intercomparison. The Okmok phoenix ash-clouds are generally exiting simulation domains with heights of 20–50 km, but the tall domain of run 18 (85 km) shows that neutral buoyancy is achieved in the upper stratosphere. Including decompression cooling might lower this level. Another factor that has the opposite effect is that the solid–gas momentum coupling is neglected below a solid volume fraction of $10^{-4}$ (the minimum value of $\varepsilon_m$ tracked is $10^{-8}$, SI Section S1). Particle mass loading at the inlet, $(1 - x_{w(v)})/x_{w(v)}$, span from 10 to $10^3$ (Table S1). In collapsing columns and PDCs, these loadings remain high (>10), whereas they are between 1 and 10 in the phoenix ash-clouds. At high loadings (≥10), mesoscale particle clusters can form, which generate volume fraction and velocity fluctuations that produce and sustain fluid turbulence (e.g., Capecelatro et al., 2015). The drag sub-grid model we used (Gidaspow 1994) does not capture the enhanced settling of such mesoscale clusters, nor the shock-related effects at high Ma, underestimating mean particle velocity (Capecelatro and Wagner 2024). We expect that including such effects would change the coupling/decoupling dynamics of the three particle sizes we simulate, possibly affecting air entrainment into the plume.

Our results show that most phoenix ash-clouds occur in pulses rather than in a continuous fashion. This is easily understandable for clouds generated from the body and head of the PDC flowing out of the caldera because of the encounters with complex topography. It is perhaps more surprising for the central plume because it is fed continuously by the convergence of the collapsing fountain atop the ring vent. The pulsatory nature of the central plume is explained by the appearance of one region above which the plume is positively buoyant and below which it is negatively buoyant (Valentine and Cole 2021). The appearance of a neutrally buoyant point is explained by a reduction of the bulk density of the ascending mixture of gas and particles caused by the interaction with the atmosphere (Bursik and Woods 1991).

An incomplete covering of the $x_{w(v)}$–$W$–$Q$ parameter space with the constraint that PDCs reach 26 km after ejection of ≥14 km³ DRE of material yielded two runs (18 and 19) compatible with field observations. The fountaining dynamics of these compatible runs is consistent with the presence of agglutinate deposits on the caldera flanks. The analysis of the runs without steady, buoyant PDC liftoff led us to propose *1)* a regime diagram (Fig. 5) constraining the $Q$ and $Ri$ values producing PDC; *2)* a relationship to determine at which critical mass eruption rate, $Q_{cr}$, the observed volume of ejecta, $V_{obs}$, reaches the observed PDC runout; *3)* how $Q_{cr}$ depends on $W$ and $x_{w(v)}$ (Fig. 6). These relationships populate the $x_{w(v)}$–$W$–$Q$ space and allow us to extrapolate our results to provide reasonable estimates of the eruptive parameters of the Okmok II eruption.

Using a reasonable $V_{obs}$ of 29 km³ DRE and that PDC hindrance by topography may delay head propagation by up to 10 minutes, we find that $Ri$=1–14 and $Q_{cr}$=1.2–3.9×10¹¹ kg/s, which encompasses the 1.4×10¹¹ kg/s estimated by Burgisser (2005). In our runs, vent pressure was assumed to be atmospheric. Estimates from modeling coupling conduit flow to plume atmospheric dispersal (Neri et al. 2002; Esposti Ongaro et al. 2006) suggest potentially higher vent pressures of 0.1–8 MPa with 0.3≤$Ma$≤9. Estimating $x_{w(v)}$ is challenging because not all the exsolved gas is used to propel the fountaining jet; a fraction of it remains within the scoria as trapped gas bubbles. Considering independent measures of Vulcanian jets and how fragmented the Okmok ejecta is (SI Section S8), we estimate that 30–50 % of the total gas fraction was propelling the jet. Taking the upper estimate of 1.7 wt% for the total water fraction, this implies that $x_{w(v)}$=0.5–0.85 wt%. Figure 12 extrapolates our results by showing $W$ and $Ma$ as a function of vent pressure for this bracket of $x_{w(v)}$ values at $Q_{cr} = 3.9 \times 10^{11}$ kg/s and $Ri$=14. It assumes no runout inertia after eruption shutdown, which we have shown to be true only for PDC hindered by distal topography. Such hindrance is consistent with the stratified deposits found at those distal locations (Fig. 1B). Figure 12 shows the severe limitations of establishing precise eruptive parameters of historical eruptions and the need for additional constraints, such as conduit flow modeling (Esposti Ongaro et al. 2006) or independent estimates of $x_{w(v)}$.

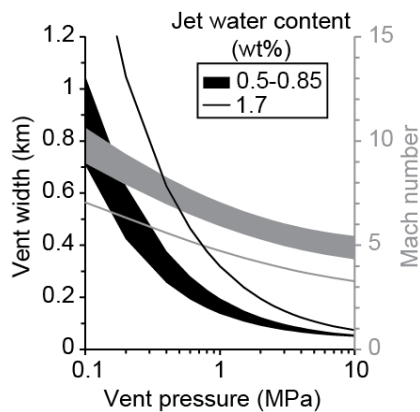

**Figure 12**: Best guess at constraining the eruptive parameters of the Okmok II eruption. Vent width ($W$, left black axis) and Mach number ($Ma$, right gray axis) as a function of vent pressure and jet water content (the areas cover the most likely range of 0.5–0.85 wt% $H_2O$ and the curves mark the upper estimate of 1.7 wt% $H_2O$). As the maximum $Q_{cr}$ was used, the $W$ values are also maximum. Conversely, lowering $Ri$ (here set to the maximum value of 14) would increase $W$. Fig. 6 shows the effects of varying $Ri$ and PDC slowdown by topography at a fixed vent pressure of 0.105 MPa.

Our simulations ignored PDC travel over water because the cross section focused on the interaction between the PDC and the series of SW hills. Towards the east, however, PDCs crossed the strait between Umnak and Unalaska islands (Fig. 1A). Overwater travelling causes particle loss, which dilutes the PDC and favors the generation of phoenix ash-clouds (Dufek and Bergantz, 2007a; Dufek and Bergantz, 2007b). While this seems an incentive to address overwater travel, a significant complexity is added by the rapid formation of pumice rafts able to support bouncing of lithics. The additional generation of offshore phoenix ash-clouds cannot thus be excluded but the inhibiting role of pumice rafts probably minimized such sources of stratospheric injection. Addressing that issue was deemed to exceed the predictive capability of our PDC modeling.

## *5.1 Estimating S stratospheric load*

Our results show that a fluctuating emission rate could increase the loading we estimated from a steady mass eruption rate at the vent, but it is unlikely. A buoyant plume carrying a significant amount of ejecta would have left recognizable fall deposits interbedded with the massive deposits, which are not seen of the field (Burgisser, 2005; Larsen et al., 2007). Signs of fall deposits would also be present within or atop the stratified deposits found on the topographic highs of Umnak, and within those blanketing Unalaska. Field observations (including the absence of erosion between eruptive phases) are thus not compatible with any significant syn-eruptive fallout episode. The liftoff of phoenix ash-clouds at the end of the eruption is not expected to leave discernible traces in the deposits because it transports material far from source, leaving the extended domain of 83 by 85 km to reach Greenland (McConnell et al. 2020; Peccia et al. 2023).

The flow path we simulated crosses a rugged part of the topography surrounding the caldera. Other reliefs, such as Idak plateau (Fig. 1A), could generate phoenix clouds at various locations, but some flow directions, such as towards the SE, lack the topography typically fostering such clouds. These considerations point to <1 wt% being a most likely upper estimate of $F_s$ syn-eruptively if no PDC steady liftoff occurs.

We thus estimate from Fig. 11 that, overall, 58–64 wt% of the emitted volcanic gas was able to reach the stratosphere, largely in post-eruptive PDC liftoff plumes. The corresponding stratospheric load of S controls the atmospheric forcing and the climatic response in temperature and precipitation changes (McConnell et al., 2020; Peccia et al. 2023). The sulfur load of the juvenile magma has been determined by the petrologic method, which quantifies the difference between the S contained in melt inclusion and that remaining in the glass, correcting for crystal content (Peccia et al. 2023). This method suggests that ~1500 ppm of sulfur was degassed during the climactic PDC phase of Okmok II. The total amount of juvenile magma emitted by this phase can be obtained from the total volume corrected for the amount of lithics. Using a total erupted volume of 29 km$^3$ DRE, Peccia et al. (2023) calculate a total juvenile magma mass of ~4.3×10$^{13}$ kg, which in turn produced a total sulfur load of ~62 Tg S. Assuming no selective precipitation of S during the vent–stratosphere travel, we can equate the stratospheric sulfur fraction to the jet gas fraction. Neglecting the volatiles remaining dissolved in the scoria is reasonable because the whole rocks and matrix glasses are degassed in sulfur (Peccia et al. 2023). Using that 30–50 % of the total gas fraction was propelling the jet (SI Section S8), we find that only 18.6–31 Tg S was liberated into the atmosphere by fragmentation. Thus, our modeling suggests that 11–20 Tg S was injected into the stratosphere. This amount of stratospheric S is lower than the 38–48 Tg S inferred previously from source location, ice core sulfate concentrations, and hemispheric depositional asymmetry (McConnell et al., 2020; Pearson et al., 2022), while it overlaps nicely with the 18–22 Tg S inferred from tree ring and speleothem temperature reconstructions and climate model response to S (Peccia et al. 2023b). The reader is directed to the work of Peccia et al. (2023b) and its Fig. 4 for an in-depth analysis of the inferences that can be drawn from these different results while noting that Peccia et al. (2023b) assumed that all the gas was propelling the jet and used the preliminary $F_g$ range of 2.5–25 wt%, which was estimated before being able to fully simulate post-eruptive dynamics (Section 4.3).

## *5.2 Study intercomparison of the buoyant–collapse transition*

We tallied the various quantities thought to influence the buoyant–collapse transition ($Q$, $x_{w(v)}$, $A$, $v_{in}$, $\varepsilon_s$, $\varepsilon_s v_{in}$, $P_n$, $Ri$, $Ma$, $Tg_m$, $K_p$, $\tilde{A}$, $\tilde{q}_b$, and $\tilde{Q}$, Section 1) in 141 runs from 10 studies performing 2D and/or 3D simulations of eruptive jets (SI Section S9). Caldera filling was left aside because it is addressed by only one study. Temperature was not considered as an independent variable as its effect is captured by $\rho_b$. 1D studies were also left out of this tally because, although dimension reduction has its own merits, the comparison with multidimensional simulations shows that 1D models are unable to faithfully reproduce the air entrainment that controls partial column collapse (Costa et al. 2016).

It was intuited early that the buoyant–collapse transition could not be captured by a single parameter (Valentine and Wohletz 1989; Dobran 1992). Figure 13 shows representative pairings of parameters that have been proposed in the literature to be the controls of the buoyant–collapse transition. Runs belonging to the same studies share a symbol shape and runs are classified according to the proportion of material that collapsed from the eruptive column. The various regime diagrams correctly discriminate eruptive regimes in the original study that proposed them. For instance, the pair $\varepsilon_s v_{in}$–$A$ (Fig 13E) was proposed by Dufek and Bergantz (2007b), and all their runs with $\varepsilon_s v_{in}$ values below 0.3–0.4 are fully buoyant with the exact $\varepsilon_s v_{in}$ value depending on vent area $A$. No pairs of variables, however, yield groups of runs according to their eruptive regime. In other words, bivariate discrimination of the eruptive regime of all runs is not possible.

Our results show that the pair $Q$–$Ri$ is able to characterize the buoyant–collapse transition at Okmok (Fig. 5). When considering this parameter space (which to our knowledge has never been proposed as a regime diagram), Fig. 13F shows that all studies but one (the pioneering study of Valentine and Wohletz, 1989) define a field of buoyant plumes and a field of (partially or fully) collapsing columns. This corresponds to an overall 70% success rate in predicting the correct eruptive regime. Even removing that study (or any single other one), no other variable pair shows such a consistency; there are always several runs with differing regimes overlapping (e.g., Fig. 13B). We did not find a robust reason why the Valentine and Wohletz (1989) runs would behave differently than those of the other studies. A possibility is that their vent (and thus the overlaying jet) is discretized by only 1–3 computational cell(s) depending on the run. By analogy with what is observed in 1D models (Costa et al. 2016), this might not have been sufficient to capture air entrainment and the ensuing partial or total collapse of the jet. Although a reanalysis of that study would surely shed light on the reasons beyond this apparent discrepancy, it is beyond the scope of this work.

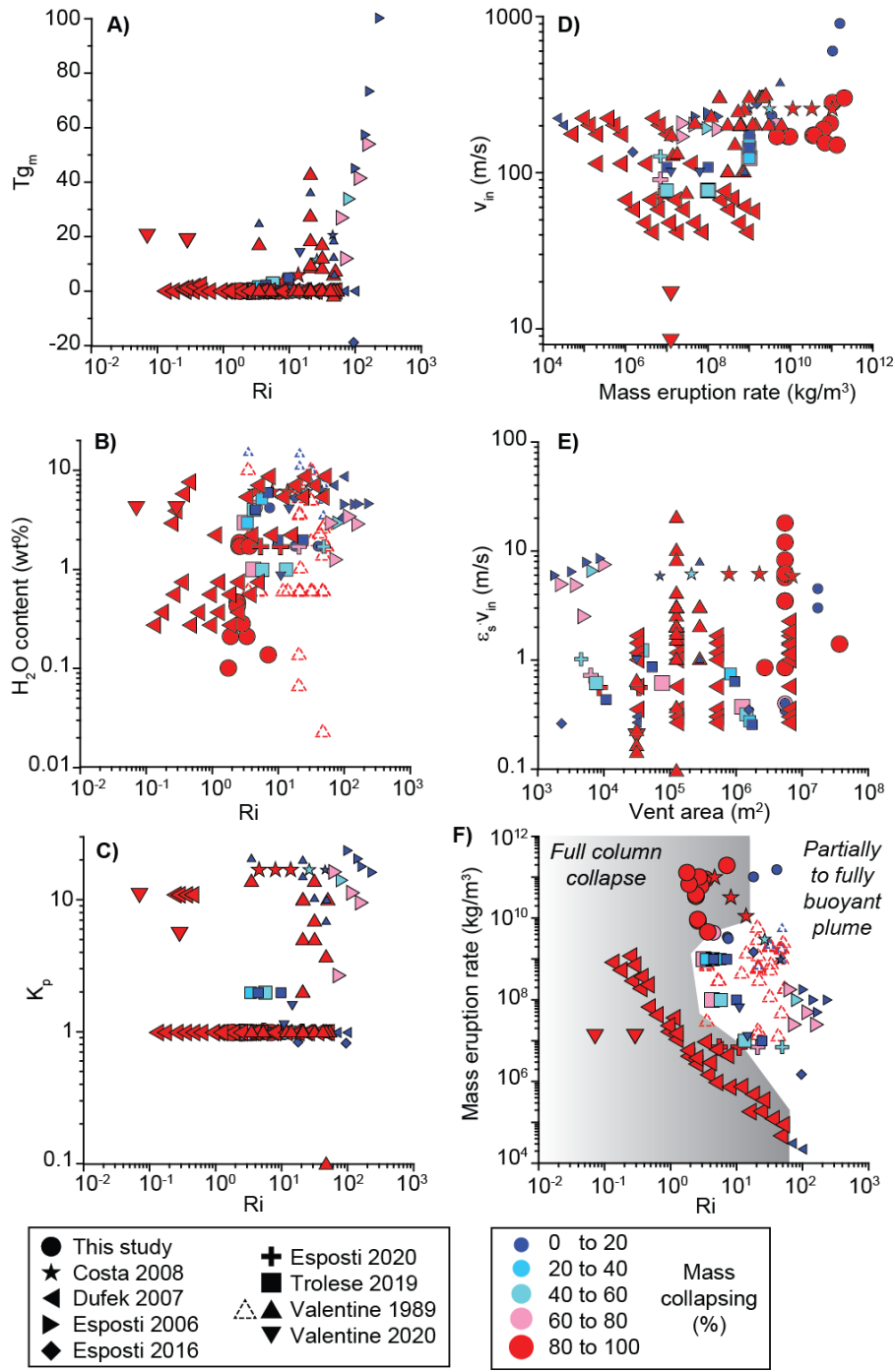

**Figure 13**: Bivariate parameter space of representative variables pairs proposed to play a major role in controlling eruption column collapse. Runs performed in a given study share symbol shapes (data and their sources are listed in SI Section S9). Symbol colors represent the fraction of mass collapsing so that 0% is a fully buoyant eruptive column feeding an umbrella cloud and 100% is a fully collapsing fountain generating pyroclastic density currents. Some triangles have been represented with dashed outlines to highlight their apparent peculiar behavior compared that of other studies. A-E) Representative selection of parameter pairs that do not discriminate the buoyant–collapse transition. F) Parameter pair that discriminates the buoyant–collapse transition for all runs of all studies but one (see text). The gray surface indicates the field of full column collapse.

Many combinations of more than two parameters, such as $Ri$–$Tg_m$,–$K_p$ (Valentine and Wohletz 1989), were not more successful in discriminating the eruptive regime (SI Section S10). The main exception is the $\tilde{A}$–$\tilde{q}_b$–$\tilde{Q}$ combination proposed by Koyaguchi and Suzuki (2018), which takes into account near-vent decompression and is based on analytical solutions of one-dimensional eruption models. Figure 14 shows the data distribution in the $\tilde{A}$–$\tilde{q}_b$–$\tilde{Q}$ space with respect to the $\tilde{Q}$ collapse criterion of Eq. (15), which has an overall success rate of 81%. Figure 14 suggests that adjusting the collapse criterion would not improve this rate much because some buoyant columns have nearly the same $\tilde{A}$–$\tilde{q}_b$–$\tilde{Q}$ values as collapsing columns. This is confirmed by the modest 1% increase of the success rate when considering the full range of values of the adjustment constant and the air entrainment coefficient that control $Ri_K$ (and thus $\tilde{Q}$, Eq. 14, see SI Section S5).

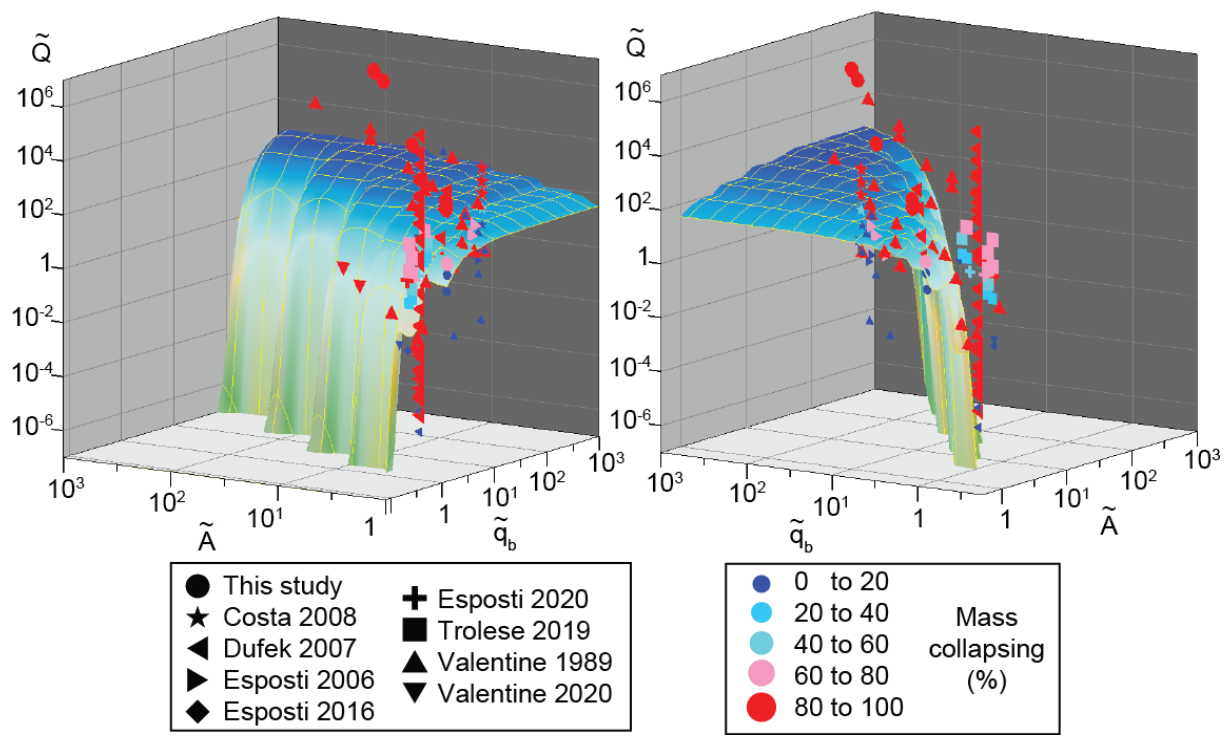

**Figure 14:** Parameter space of the three dimensionless variables proposed by Koyaguchi and Suzuki (2018) to play a major role in controlling eruption column collapse: the normalized ratio of the cross-sectional areas at the crater top and base ($\tilde{A}$), the normalized magma supply rate at the crater base ($\tilde{q}_b$), and the normalized mass eruption rate ($\tilde{Q}$). Left and right diagrams show two spatial combinations of horizontal axes to help visualization. Runs performed in a given study share symbol shapes (data and their sources are listed in SI Section S9). Symbol colors represent the fraction of mass collapsing (0% is a fully buoyant column and 100% is a fully collapsing fountain). The surface of critical $\tilde{Q}$ values

was calculated by building a grid of *Ma* and $K_t$ values that were converted into $\tilde{A}$ and $\tilde{q}_b$ values using their respective definitions. Data points above this surface are predicted to yield collapsing columns (i.e. red, pink, or possibly cyan symbols). Some collapsing columns yielding very high $\tilde{Q}$ and $\tilde{q}_b$ values (such as $10^{45}$) were clipped from view.

In pressure-balanced inlet conditions, such as in our runs, $K_t \sim 1$. Using Eq. (11), the Koyaguchi and Suzuki (2018) collapse condition becomes $Q \geq Ri_K^2 Ri^2 A\rho_b v_{in}$ in dimensional units. Using the definition of $Q$ (Eq. (9)), we obtain a collapse criterion, $Ri \leq Ri_K^{-1}$, which depends only on the Richardson number. It has an 88% success rate with our runs (94% with the full $\tilde{Q}$ criterion). We thus suggest that the *Q–Ri* regime diagram is able to capture the first-order bivariate layout of the buoyant–collapse transition. Unquestionably, the multivariate $\tilde{Q}$ criterion of Koyaguchi and Suzuki (2018) is more general and thus accommodates a much wider range of jet initial conditions. The source Richardson number is the ratio between kinetic energy and buoyancy at the vent; jets with high *Ri* are more likely to have enough kinetic energy to rise and feed a buoyant plume than to collapse back to the ground. The mass eruption rate is a dimensional quantity sensitive to the compound effects of jet size, bulk density, and speed, which respectively control how far from positively buoyant the jet is (density) and how easily air can be engulfed into the jet (speed and size). Figure 13F and the simplified $Ri_K$ criterion suggest that air entrainment compensates for the *Ri* effect between $10^7$ and $10^{10}$ kg/s. In other words, pressure-balanced jets need less kinetic energy to become buoyant within this bracket of *Q* values than outside of it because air entrainment is optimized within this bracket.

## 6. Conclusions

This work presents results from computational fluid dynamics aimed at estimating the amount of material reaching the stratosphere during the eruption of Okmok II. Axisymmetric, 2D simulations were carried out with the two-phase flow model MFIX-TFM. The dynamics of the climactic phase of the eruption were modelled with several combinations of mass eruption rate, jet water content, vent size, particle size and density, topography, and emission duration. Model outputs were constrained by the volume of the deposits and their distribution on the SW of the caldera, where four successive hills caused deposit thickness to vanish progressively. Stratospheric gas injection was quantified by mass balance.

Our simulations suggest that an average mass eruption rate of 1.2–3.9×10$^{11}$ kg/s is consistent with field observations and an erupted volume of 29 km$^3$ DRE. The exact value within this range depends on whether PDCs are temporarily blocked by topography, with higher values corresponding to unhindered PDCs. If there is no steady, buoyant liftoff of the PDC during eruption, this rate can be achieved by the combinations of vent width, jet water content, and vent pressure depicted in Fig. 12.

The parameter space defined by the source Richardson number and the mass eruption rate is able to characterize the buoyant–collapse transition at Okmok. We extended this result to 141 runs from 10 studies performing 2D and/or 3D simulations of eruptive jets to show that such $Q$–$Ri$ regime diagram (Fig. 13F) is able to capture the first-order layout of the buoyant–collapse transition in all studies except one (i.e. an overall 70% success rate). The highest degree of successful discrimination, 82%, is achieved by the multivariate $\tilde{Q}$ criterion of Koyaguchi and Suzuki (2018) that takes a wider range of jet initial conditions into account (Fig. 14).

During eruption, stratospheric (>11 km high) injections occur in pulses rather than in a continuous fashion. The first injection is caused by the convergence of the collapsing fountain atop the ring vent in the center of the caldera that generates a central plume. This central plume is able to inject some gas above the tropopause before collapsing back on itself, stopping the stratospheric injection. The next pulses reaching the stratosphere are successive phoenix ash-clouds issued from the head and main body of the PDC as it encounters complex topography. In conditions compatible with field observations, all these syn-eruptive injections add up to <1 wt% of gas in the stratosphere. If there is a steady, buoyant liftoff of the PDC during eruption, the amount of stratospheric gas rises to <10 wt%. The end of the eruption causes a buoyant liftoff of the phoenix ash-clouds from the most dilute parts of the PDC, which can dominate the stratospheric injection of other eruptive phases, with up to 58–64 wt% injected gas fraction (with respect to the total volume erupted) if the mass eruption rate at the vent is steady.

A fluctuating emission rate generating a transient buoyant plume or an efficient final liftoff due to seawater interaction could increase the loading we estimated for a steady mass eruption rate, but modeling and field constraints suggest that it is unlikely. The strongest field constraint is that a sustained buoyant plume would have left recognizable (and yet absent) fall deposits interbedded with the massive PDC deposits.

The stratospheric load of S controls the atmospheric forcing and the subsequent climatic response in temperature and precipitation changes. Using a total erupted volume of 29 km$^3$ DRE and petrological constraints on the amount of S degassed, our study suggests that Okmok II injected 11–20 Tg S into the stratosphere. This amount is consistent with the observed ice sheet deposition, paleoenvironmental estimates of cooling, and modeling of climate sensitivity to sulfur loads (Peccia et al. 2023).

## Acknowledgments

We thank G. Valentine for helping us identifying gargle dynamics in our simulations and J. Larsen for suggesting exploring alternate pre-eruptive topography. We appreciated the thoughtful comments of C. Vidal, E. Breard and one anonymous reviewer and the comments and editorial handling of T. Esposti Ongaro. Their collective inputs strengthened the manuscript. We would like to thank Eric Breard and Jordan Musser for sharing their MFIX implementation of the gas cooling decompression term.

## Funding

This project was partially funded by grant ANR-19-0315CE-0007 from Agence Nationale pour la Recherche. The computations presented in this paper were performed using the GRICAD infrastructure (gricad.univ-grenoble-alpes.fr), which is supported by Grenoble research communities.

# Statements and Declarations

No ethical question is raised by this study. All authors and co-authors have given their consent to be included in this manuscript and to publish it. The authors declare no competing interests.

**Table 1:** Run parameters and inlet conditions. $M$ is the number of solid phases, $\varepsilon_s$ is the total particle volume fraction, $x_{w(v)}$ is the gas mass fraction, $\rho_b$ is the bulk density, $v_{in}$ is the inlet velocity, $Ma$ is the Mach number, $W$ is the vent width, $Q$ is the mass eruption rate of the solids, $t_Q$ is the emission duration, $t_{end}$ is the simulation end, $V_s$ is the cumulative volume of erupted solids at $t_{end}$ (DRE stands for Dense Rock Equivalent), BC is the nature of the distal vertical boundary (F = free outflow/inflow, W = free slip wall), $F_g$ is the gas fraction that reaches the stratosphere at $t_{end}$, $L_D$ is the domain length, and $H_D$ is the domain height. Runs with the suffix "stop" were run from 0 to $t_Q$ with the inlet conditions of the run with the same prefix and then continued until $t_{end}$ with the inlet turned off.

| Run | $M$ | $\varepsilon_s$ (vol%) | $x_{w(v)}$ [a] (wt%) | $\rho_b$ [a] (kg/m³) | $v_{in}$ (m/s) | $Ma$ [a] | $W$ (m) | $Q$ [a] (kg/s) | $t_Q$ (s) | $t_{end}$ (s) | $V_s$ [a] (km³) | $V_s$ DRE [a] (km³) | BC [b] | $F_g$ (wt%) | $L_D$ (km) | $H_D$ (km) |
|---|---|---|---|---|---|---|---|---|---|---|---|---|---|---|---|---|
| 3 | 1 | 0.5 | 1.7 | 10.2 | 900 | 8.9 | 600 | 1.6×10¹¹ | 80 | 80 | 6.2 | 5.0 | F | 82 | 83 | 51 |
| 4 | 1 | 0.5 | 1.7 | 10.2 | 600 | 5.9 | 600 | 1.0×10¹¹ | 114 | 114 | 5.9 | 4.7 | F | 72 | 83 | 51 |
| 5 | 3 | 2 | 0.5 | 36.8 | 173 | 3.3 | 200 | 3.5×10¹⁰ | 638 | 638 | 12.2 | 9.0 | **F+W** | 8.0 | 43 | 20 |
| 5stop | | | | | | | | | 310 | 566 | 5.9 | 4.4 | **W** | 15 | 43 | 20 |
| 6 | 3 | 0.15 | 6.0 | 2.9 | 231 | 1.2 | 200 | 3.5×10⁹ | 400 | 400 | 0.8 | 0.6 | W | 51 | 21 | 20 |
| 7a [c] | 3 | 0.5 | 1.9 | 9.3 | 170 | 1.6 | 200 | 8.6×10⁹ | 100 | 100 | 0.5 | 0.3 | F | 0 | 21 | 20 |
| 7b [c] | 3 | 0.15 | 6.0 | 2.9 | 231 | 1.2 | 200 | 3.5×10⁹ | 300 | 300 | 0.9 | 0.6 | F | 18 | 21 | 20 |
| 7c [c] | 3 | 0.5 | 1.9 | 9.3 | 170 | 1.6 | 200 | 8.6×10⁹ | 463 | 463 | 1.6 | 1.2 | F | 30 | 21 | 20 |
| 8 | 1 | 4 | 0.2 | 80.2 | 206 | 5.8 | 200 | 9.1×10¹⁰ | 280 | 280 | 12.8 | 10.2 | F | 0.19 | 43 | 20 |
| 8stop | | | | | | | | | 165 | 434 | 7.5 | 6.0 | F | 8.9 | 43 | 20 |
| 9 | 1 | 0.5 | 1.7 | 10.2 | 170 | 1.7 | 200 | 9.4×10⁹ | 646 | 646 | 3.0 | 2.4 | W | 29 | 43 | 20 |
| 9stop | | | | | | | | | 394 | 1365 | 1.9 | 1.5 | W | 63 | 83 | 20 |
| 10 | 1 | 3 | 0.3 | 60.2 | 190 | 4.6 | 200 | 6.3×10¹⁰ | 308 | 308 | 9.7 | 7.7 | F | 0.27 | 29 | 20 |
| 11 | 1 | 0.5 | 1.7 | 10.2 | 280 | 2.8 | 1200 | 1.0×10¹¹ | 400 | 400 | 20.2 | 16.6 | F | 7.8 | 83 | 51 |
| 11stop | | | | | | | | | 160 | 1307 | 8.3 | 6.6 | F | 59 | 83 | 51 |
| 12 | 3 | 2 | 0.5 | 40.2 | 173 | 3.4 | 200 | 3.5×10¹⁰ | 369 | 369 | 7.1 | 5.7 | **F** | 3.6 | 43 | 20 |
| 13 | 1 | 2 | 0.4 | 40.2 | 173 | 3.4 | 200 | 3.8×10¹⁰ | 423 | 423 | 8.1 | 6.5 | **F** | 4.3 | 43 | 20 |
| 14 | 1 | 4 | 0.2 | 80.2 | 155 | 4.4 | 200 | 6.9×10¹⁰ | 279 | 279 | 9.6 | 7.7 | F | 0.66 | 29 | 20 |
| 14stop | | | | | | | | | 180 | 304 | 6.2 | 4.9 | **F** | 1.1 | 29 | 20 |
| 15 | 1 | 0.2 | 4.2 | 4.2 | 200 | 1.3 | 200 | 4.4×10⁹ | 315 | 315 | 0.7 | 0.6 | F | 35 | 29 | 20 |
| 16 | 1 | 0.2 | 4.2 | 3.1 | 200 | 1.3 | 200 | 3.2×10⁹ | 308 | 308 | 0.7 | 0.5 | F | 42 | 29 | 20 |
| 17 | 1 | 0.5 | 1.7 | 10.2 | 170 | 1.7 | 100 | 4.7×10⁹ | 740 | 740 | 1.7 | 1.4 | F+W | 28 | 43 | 20 |
| 17stop | | | | | | | | | 484 | 1217 | 1.1 | 0.9 | W | 58 | 43 | 20 |
| 18 | 1 | 8 | 0.1 | 160.2 | 150 | 6.1 | 200 | 1.3×10¹¹ | 600 | 600 | 39.8 | 31.8 | F | 9.7 | 83 | 85 |
| 18stop[d] | | | | | | | | | 208 | 1450 | 14.1 | 11.3 | F | 58 | 83 | 85 |
| 19 | 1 | 6 | 0.1 | 120.2 | 300 | 10.5 | 200 | 2.0×10¹¹ | 218 | 218 | 21.7 | 17.4 | F | 0.18 | 83 | 51 |
| 19stop[e] | | | | | | | | | 178 | 1924 | 18.0 | 14.4 | W | 55 | 83 | 51 |

[a] dependent variables. The other variables are independent.
[b] when two boundary types are listed, the run started with the free flow boundary and was continued with a distal wall. The font indicates whether the solid phase(s) could be in close packing conditions, in which case a solid pressure based on plastic flow theory is calculated (bold font=close packing possible, normal font=no close packing)
[c] run comprising 3 consecutive parts (a, b, and c) with changing inlet conditions. The total run time for run 7 is thus $t_{end}$=863 s.
[d] the inlet speed was decreased to 100 m/s from 208 to 215 s before being set to 0.
[e] the inlet speed was decreased to 200 m/s from 177 to 185 s before being set to 0.

# Appendix

| Symbol | Description, unit |
| --- | --- |
| $A$ | Vent area, m$^2$ |
| $\tilde{A}$ | Normalized ratio of the areas at the crater top and base |
| $C_{pa}$, $C_{pw}$, $C_{pg}$, $C_{pm}$ | Heat capacity (air, water, gas, solid phase $m$), J kg$^{-1}$ K$^{-1}$ |
| $\mathbf{D}_g$ | Rate of strain tensor, 1/s |
| $d_m$ | Particle size of solid phase $m$, m |
| $f_d$ | Drag coefficient, kg m$^{-3}$ s$^{-1}$ |
| $F_g$ | Gas weight fraction that reaches the stratosphere |
| $\mathbf{g}$ | Gravity acceleration vector, m/s$^2$ |
| $H_D$ | Domain height, m |
| $K_p$ | Vent pressure over atmospheric pressure |
| $k_t$ | Turbulent kinetic energy, J/kg |
| $L_D$ | Domain length, m |
| $m$, $M$ | Solid phase index, number of solid phases |
| $Ma$ | Mach number |
| $M_{w(in)}$, $M_{w(out)}$ | Mass of water crossing a level (vent, tropopause), kg |
| $Nu_m$ | Nusselt number of solid phase $m$ |
| $P_g$, $P_a$, $P_v$ | Pressure (gas, heated atmosphere at inlet, gas at inlet), Pa |
| $P_n$ | Rouse number |
| $Q$, $Q_{cr}$ | Mass eruption rate of solids (standard, critical), kg/s |
| $\tilde{Q}$ | Normalized mass eruption rate |
| $\tilde{q}_b$ | Normalized magma supply rate at the crater base |
| $r_1$, $r_2$, $R_{eq}$ | Vent radius (inner, outer, equivalent), m |
| $Ri$, $Ri_{cr}$, $Ri_K$ | Richardson number (source, critical for $Q_{cr}$, critical for $\tilde{Q}$) |
| $S$ | Advected scalar |
| $S_g$, $S_m$ | Total stress tensor (gas, solid phase $m$), Pa |
| $t$, $t_{end}$ | Time (variable, end of simulation), s |
| $T_g$, $T_m$ | Temperature (gas, solid phase $m$), K |
| $Tg_m$ | Jet driving force over jet buoyancy force |
| $t_Q$ | Emission duration, s |
| $\mathbf{v}_g$, $\mathbf{v}_m$ | Velocity (gas, solid phase $m$), m/s |
| $v_{in}$ | Inlet speed, m/s |
| $V_{obs}$, $V_s$ | Erupted volume (observed, at $t_{end}$), m$^3$ |
| $W$ | Vent width, m |
| $x_{w(v)}$ | Gas mass fraction at the inlet |
| $\gamma_m$ | Heat transfer between solid phase $m$ and gas, W m$^{-3}$ K$^{-1}$ |
| $\varepsilon_g$, $\varepsilon_m$, $\varepsilon_s$ | Volume fraction (gas, solid phase $m$, total solids) |
| $\varepsilon_t$ | Dissipation rate of turbulent kinetic energy, W/kg |
| $\kappa_g$, $\kappa_m$ | Heat conductivity (gas, solid phase $m$), W K$^{-1}$ m$^{-1}$ |
| $\rho_a$, $\rho_w$, $\rho_g$, $\rho_m$, $\rho_b$ | Density (air, water, gas, solid phase $m$, bulk), kg/m$^3$ |
| $\boldsymbol{\tau}_g$ | Gas stress tensor, Pa |
| $\mu_g$, $\mu_{gt}$ | Gas viscosity (molecular, turbulent), Pa s |